\documentclass[12pt]{amsart}
\usepackage{amssymb,amsmath}
\usepackage{latexsym}
\usepackage{graphicx}
\usepackage{geometry}                % See geometry.pdf to learn the layout options. There are lots.
\usepackage{color}

\newcommand{\R}{\mathbb R}

\newcommand{\ket}[1]{|\kern.3ex#1\kern.3ex\rangle}
\newcommand{\bra}[1]{\langle\kern.3ex #1 \kern.3ex|}
\newcommand{\scalar}[2]{\langle\kern.3ex #1 \kern.3ex|\kern.3ex#2\kern.3ex\rangle}
\newcommand{\norm}[1]{\|\kern.3ex#1\kern.3ex \|}

\def\lg{\langle }
\def\rg{\rangle }

\def\lu{\mbox{\large 1}}
\def\ud{\mathrm{d}}
\def\mq{\mathfrak{q}}
\def\mp{\mathfrak{p}}
\def\ap{a_P}
\def\q{\tilde{q}}
\def\p{\tilde{p}}
\def\qm{\tilde{\mathfrak{q}}}
\def\pm{\tilde{\mathfrak{p}}}

\begin{document}
\title[Wavelet Quantum Cosmology]{Smooth big bounce from affine quantization}

\author[H Bergeron, A Dapor, J-P Gazeau  and P Ma\l kiewicz]{Herv\'e Bergeron, Andrea Dapor, Jean Pierre Gazeau  and Przemys\l aw Ma\l kiewicz}
\address{ISMO, UMR 8214 CNRS, Univ Paris-Sud,  France
} \email{herve.bergeron@u-psud.fr}

\address{APC, UMR 7164 CNRS, Univ Paris  Diderot, Sorbonne Paris Cit\'e, 75205 Paris, France}\email{gazeau@apc.univ-paris7.fr}

\address{Centro Brasileiro de Pesquisas F\'{\i}sicas (CBPF), 
22290-180 Rio de Janeiro, 
Brasil
} 
\email{gazeaujp@cbpf.br}

\address{National Centre for Nuclear Research, Ho\.za 69, 00-681 Warsaw, Poland} \email{Przemyslaw.Malkiewicz@fuw.edu.pl}

\address{University of Warsaw, Ho\.za 69, 00-681 Warsaw, Poland} \email{adapor@fuw.edu.pl}

%\thanks{
%The work was supported by Polonium program No. %27690XK {\it Gravity and Quantum Cosmology}.}

\date{\today}

\begin{abstract}
We examine the possibility of dealing with gravitational singularities on a quantum level through the use of coherent state or wavelet quantization instead of canonical quantization.   We consider the Robertson-Walker metric coupled to a perfect fluid. It is the simplest model of a gravitational collapse and the results obtained here may serve as a useful starting point for more complex investigations in future. We follow a quantization procedure based on affine coherent states or wavelets built from the unitary irreducible representation  of the affine group of the real line with positive dilation.  The main issue of our approach is the appearance of a quantum centrifugal potential  allowing for regularization of the singularity, essential self-adjointness of the Hamiltonian, and  unambiguous quantum dynamical evolution.
\end{abstract}



\maketitle
\tableofcontents
\section{Introduction}
The purpose of this article is to examine the implementing of the coherent state quantization, as presented for instance in Chapter 11 of the recent \cite{G2000},  in the study of the gravitational singularities. We consider the Robertson-Walker metric coupled to a perfect fluid. It is the simplest model of a gravitational collapse and the results obtained here may be a useful starting point for more complex investigations in future. In particular, the examination of such issues as the probability for inflation in the presence of a scalar field in quadratic potential (see e.g. \cite{Vilenkin}) are postponed until next papers.

Canonical, or Weyl, or Weyl-Wigner, quantization of a Friedman-Lemaitre universe with an eye towards the fate of gravitational singularity has been studied extensively. Early treatments include Blyth and Isham \cite{ish}, featuring a discussion on the ambiguous meaning of singularity resolution, and the work by Lapchinskii and Rubakov \cite{rub}, who obtained a quantum non-singular perfect fluid-filled universe. However, the interpretation of these and other, more recent, results is not obvious for at least three reasons. Firstly, in \cite{rub} it was shown that the classical and singular evolution can be replaced by a unitary and thus non-singular one, provided one fixes an appropriate boundary condition to ensure self-adjointness of the Hamiltonian. Unfortunately, there are infinitely many `equally good' choices for the boundary condition and the choice has to be made without a clear justification. Furthermore, the singularity resolution corresponds simply to a reflection of the wave function against the singularity, while one would expect the quantum effects to appear and play a role in dynamics already before the singularity is reached. Secondly, at the fundamental level, the prevailing attitude towards quantum gravity is that one needs some additional input, let it be loops, strings or triangulations. The inclusion of these basic premises should give rise to novel effects in the context of mini-superspaces. Finally, there is the so-called `problem of time': in canonical quantum gravity one is forced to describe the evolution of the space with respect to a chosen degree of freedom and it may be shown that the resultant quantum theory significantly depends on this choice \cite{PM}. In this light, the expectation that Weyl quantization of mini-superspaces should lead to some important clues about the singularity may seem doubtful.

Recently, it has been argued that quantum cosmology may become an empirical science but only if developed as an effective theory \cite{bojowald}. There are simply too many unknowns, including unresolved conceptual issues and technical complexities, to have a hope that a fundamental and rigorous approach is feasible. Bojowald suggests that a good cosmological theory should therefore be flexible enough to parameterize our ignorance. Canonical  quantization is in a sense rigid and will not provide us with parametrizable physical models. For a clear and comprehensive review of various quantization methods (e.g., canonical, geometric, deformation, ...) we refer to Ali and Engli\v{s} in \cite{csenglis}. In what follows, we propose to relax the usual, canonical, quantization prescription by implementing coherent state quantization in the study of the cosmological singularity. 

The coherent state quantization was demonstrated to be a valid alternative to canonical quantization in dealing with various simple systems (see \cite{G2000,csbergeron10_12,csbergeron13}, particularly \cite{csgazeau}  for deep probabilistic aspects of the procedure, and references therein). It is a flexible method, because it allows for a reasonable amount of freedom due to the free choice of coherent states (or `wavelet basis') which determine the quantum realm of a model. In CS quantization one does not require that the Poisson bracket of basic variables is strictly mapped into the corresponding Lie algebra of the operators. Besides the standard linearity, identity correspondence $1\mapsto I$ and the self-adjointness of quantum observables, the only minimal conditions are: (i) the quantization must agree with available measurements and (ii) it should consistently admit the classical limit. In this way one allows for many more possibilities. This seems to be a desired quality for quantizing the gravitational field. Surprisingly, relaxing some of the usual quantization constraints leads to the occurrence of a quantum repulsive potential, which regularizes the singularity and leads to a unique unitary evolution across the bounce. This is the central result of this work. On top of that, CS quantization naturally provides a semiclassical counterpart for the quantum `true' Hamiltonian through the so-called `lower symbol'. Last but not least, CS have the advantage of being well-suited to deal with non-standard (e.g. non-polynomial) Hamiltonians occurring in gravitational systems, which may be useful in future investigations.

Let us emphasize that it is not an ad hoc, fine-tuned, ordering of basic operators that leads to the regularization of singularity. Rather, having given up the usual Weyl-Wigner quantization, we follow an alternative and equally general prescription based on coherent states. Since the cosmological model considered herein is defined in the phase space, which is the half-plane, and since the half-plane \underline{is} also the affine group, it is natural (and straightforward) to employ the affine group, and not the Weyl-Heisenberg group,  for the construction of coherent states. Actually, it turns out that the same mechanism of the singularity avoidance can also be derived with Weyl-Heisenberg coherent states, as is shown in Appendix \ref{D}. However, the derivation is less immediate.

The time problem understood as the ambiguity in the choice of time function is beyond the scope of the present paper. We choose to introduce perfect fluids into the model and use them to `gauge' the evolution of the remaining variables. The advantage of this approach is twofold. First, we obtain a time-independent true hamiltonian. This property simplifies the subsequent analysis and can be readily extended to anisotropic models, including the Bianchi IX type\footnote{Contrary to massless scalar fields, which are often employed in quantum cosmology, perfect fluids generally do not suppress  the oscillatory nature of the singularity in Bianchi IX model.}. Second, with this choice of time all the geometrical quantities are `physical' and thus, we may proceed with quantization of the symmetry-reduced ADM phase space with its characteristic ranges of canonical coordinates.

This paper is organized as follows. In Section 2 we recall briefly the equations of the FLRW model, we define the Hamiltonian and the relevant physical quantities. In Section 3 we present the details of our covariant quantization scheme based on an Unitary Irreducible Representation (UIR) of the affine group. At this stage some important parameters of the representation are not specified. The expressions of quantized observables are given in Section 4. In Section 5 the parameters of the representation are specified, we obtain analytical expressions, and we perform numerical simulations.  We analyze the physical aspects of our results in Section 6 and we conclude in Section 7. For connections of our approach with the affine quantization of gravity, see Appendix \ref{A}.
Appendix \ref{B} is a short review of the affine group  and its representation(s). Appendix \ref{C} introduces affine coherent states in full generality. The main result of the present paper, namely the appearance on the quantum level of the centrifugal potential and its regularizing r\^ole, is re-derived with Weyl-Heisenberg coherent states in Appendix \ref{D}.
\section{The FLRW model}

\subsection{Hamiltonian constraint}
The space + time split in the Hilbert-Einstein action leads to the Hamiltonian density as a sum of first-class constraints \cite{adm}:
\begin{equation}\label{con1}
H=NC^0+N_iC^i\, ,
\end{equation}
where
\begin{equation}\label{con2}
C^0=-q^{1/2}\left(~^3R+q^{-1}\left\lbrack\frac{1}{2}(p^k_k)^2-p^{ij}p_{ij}\right\rbrack\right),~~C^i=-2p^{ij}_{~;j}\, .
\end{equation}
Generalized variables $q_{ij}$ correspond to the three-metric on spatial sections of space-time, $~^3R$ is the three-metric Ricci scalar, and $p^{ij}$ are generalized momenta (associated with extrinsic curvature of the three-hypersurfaces) equipped with the Poisson structure
\begin{equation}\label{poisson}
\left\{q_{ij}(x),p^{kl}(x')\right\}=\delta_{(i}^{~k}\delta_{j)}^{~l}\delta(x-x')\,.
\end{equation} 
Scalar $N$ and covector $N_i$ encode the remaining components of space-time metric, which here play a role of Lagrange multipliers. In the above we have put 
\begin{equation}
\label{kappa1}
\kappa:=\frac{16\pi G}{c^3}\approx 1.24 \times 10^{-34} \,\mathrm{s}\,\mathrm{kg}^{-1}\equiv  1\, . 
\end{equation}
The units will be restored if necessary within the  quantum framework when we substitute $H\mapsto (1/\kappa)\hat{H}$ while inserting the Planck area-like  $a_P=\kappa\ap\approx 8.2 \times 10^{-68}\,\mathrm{m}^2$ instead of $\ap$.  This $a_p$ yields the natural length standard $\sqrt{a_P}$.

In the FLRW universe the line element is given by:
\begin{equation}\label{metric}
ds^2=-N(t)^2dt^2+a(t)^2\delta_{ij}\omega^{i}\omega^{j}\,,
\end{equation}
where 1-forms $\omega^i$ are invariant with respect to the homogeneity group of spatial section. The specific form of $\omega^i$ depends on the spatial curvature (see e.g. \cite{ryan}). 

For metric tensor's components in (\ref{metric}) we introduce:
\begin{equation}\label{bv}
\tilde{a}:=\frac{\int a\ud\omega}{(\int \ud\omega)^{\frac{2}{3}}},~~~ \tilde{p}_a:=\frac{-12}{(\int d\omega)^{\frac{1}{3}}}\,\int a\frac{\dot{a}}{N}\ud\omega, ~~~\{\tilde{a}, \tilde{p}_a\}=1,~~~\tilde{a}>0\,.
\end{equation}
where $\ud\omega=\omega^1\wedge\omega^2\wedge\omega^3$. The integration is performed over the whole universe if it is compact and over any finite patch otherwise. Thanks to the definition of basic variables (\ref{bv}) we get rid of the Dirac delta featuring in (\ref{poisson}), identify the three non-zero metric components, give $\tilde{a}$ the interpretation of a length and make it convenient to express the (integral form of) Hamiltonian in terms of them.

The vacuum formulation (\ref{con1}, \ref{con2}) was extended to include perfect fluids by Schutz \cite{schutz}. He used the potential-velocity formulation: $
u_{\nu}=\frac{1}{h_{\mathrm{al}}}(\phi_{,\nu}+\alpha\beta_{,\nu}+\theta s_{,\nu})$, 
where $u_{\nu}$ are the components of the proper velocity of the fluid's element, $(\phi,\alpha,\beta, \theta, s)$ are independent potentials and $h_{\mathrm{al}}$ is the specific enthalpy, which is fixed via $u_{\nu}u^{\nu}=-1$. The specification of the indicated framework to the FLRW models filled with barotropic fluid subject to the equation of state $\mathrm{p}=w\rho$ ($w = const$) is straightforward and may be found in \cite{rub}. In this case the Hamiltonian constraint is (see e.g. \cite{ham}):
\begin{equation}\label{1ham}
H=N\left(-\frac{\tilde{p}_a^2}{24\tilde{a}}-6\tilde{k}\tilde{a}+\frac{p_T}{\tilde{a}^{3w}}\right)\approx 0\,.
\end{equation}
where the dimensionless $\tilde{k}=(\int \ud\omega)^{2/3}k$ and $k=0,-1$ or $1$ depending on whether the universe is flat, open or closed\footnote{Note that the value of $\int \ud\omega$ is not arbitrary for `$a$' has the geometrical meaning of the radius of curvature. For example in the case of the 3-sphere (k=1) we have $\int \ud\omega=2\pi^2$.}. Parameters $(T, p_T)\in \R\times\R^{\ast}_+$ are a canonical pair associated with the fluid. More specifically, the meaning of $p_T$ is the fluid's energy times the volume in which it appears at the power  $w$ and $T$ is an auxiliary function of physical dimension of length at the power $1-3w$. The dimensions will explicitly agree in (\ref{1ham}) once we restore $\frac{1}{\kappa}$ in front of the gravitational part of the Hamiltonian constraint. In order to bring (\ref{1ham}) to more convenient form we define  new basic variables for geometric observables: 
\begin{equation}\label{nbv}
(q, p) := \left(\tilde{a}^{3(1-w)/2},~\frac{2}{3(1-w)} \tilde{p}_a \tilde{a}^{(3w-1)/2}\right),~~~\{q,p\}=1,~~~q>0\,.
\end{equation}
Note that the physical dimensions of $q$ and $p$:
\begin{equation}
\label{dimqp}
[q] = \mathrm{L}^{3(1-w)/2}\,, \quad [p] = \mathrm{L}^{(3w+1)/2}\, ,
\end{equation}
so that $[qp] = \mathrm{L}^2$.
Thus, after specifying the lapse $N=q^{2w/(1-w)}$,  we eventually obtain:
\begin{equation}\label{Ham}
H=\left(-\alpha(w)p^2-6\tilde{k}q^{\mu(w)}+p_T\right)\approx 0\, , 
\end{equation}
where 
$$\mu(w):=\frac{2(3w+1)}{3(1-w)}\, ,\ \alpha(w):=\frac{3(1-w)^2}{32}\, ,$$
and $[H]=L^{(3w+1)}$.

\subsection{Reduced phase space}
The Hamiltonian formulation of general relativistic models introduces first-class constraints, given in (\ref{con2}), which reflect the coordinate freedom in the Einstein theory. Here, the {\it physical} symmetry of the FLRW models enables one to make use of the preferred foliation of space-time and reduce the formulation considerably to a single constraint (\ref{Ham}). To implement quantization one may employ either the Dirac approach (`first quantize, then solve constraints') or the reduced phase space approach (`first solve constraints, then quantize'). However, both the constrained and the reduced phase space will accommodate an incomplete dynamical flow due to the singularity (in a suitable choice of time variable).\footnote{One may argue that even for non-singular models there may exist a choice of clock with an incomplete Hamiltonian flow. Here we exclude such clocks. For a discussion of the dependence of quantum theory on the choice of time (so-called `multiple choice problem') we refer reader to \cite{PM}.} In what follows we restrict ourselves to the reduced (unconstrained) phase space analysis, with the choice of the variable `$T$' as a clock. 
 
The reduction of the model goes this way: we start with reducing the symplectic form $\Omega$ to a closed two-form $\Omega_R$:
\begin{equation}
\Omega=dq\wedge dp+dT\wedge dp_T~~\rightarrow~~\Omega_R=dq\wedge dp+dT\wedge d\left(\alpha(w)p^2+6\tilde{k}q^{\mu(w)}\right)\,. 
\end{equation}
where we solved the constraint for $p_T$. The form $\Omega_R$ lives on the constraint surface and is not symplectic, because it is degenerate with the Hamiltonian vector field $v_H=\{\cdot,H\}$ being in its null direction. In order to get the physical Poisson bracket we put $T=const$ and then invert $\Omega_R$. Now our system, reduced and no longer constrained, is given by:
\begin{equation}
\label{frweqs}
\{q,p\}=1,~~h_T=\alpha(w)p^2+6\tilde{k}q^{\mu(w)},~~q>0\, . 
\end{equation}
Hence we model a singular universe as a particle moving on the half-line, where the end-point of the half-line signals singularity at which the classical dynamics terminates. From the explicit form of the Hamiltonian it is apparent that the clock $T$ is slow-gauge\footnote{A clock is slow-gauge if the singularity is reached within a finite time interval \cite{gotay}.}. Two cases are particularly interesting: (i) $w=1/3$, which corresponds to the radiation as the content of universe; (ii) $k=0$, which corresponds to flat FLRW universe, which apparently can be modeled as a freely moving particle.

The compound geometric observables of physical interest include the volume $V$ and the expansion rate $\theta$, i.e., the trace of extrinsic curvature:
\begin{equation}
\label{volrate}
V:=\int a^3\ud\omega=q^{\tfrac{2}{1-w}},~~ \theta:= \frac{2}{1-w}\frac{\dot{q}}{Nq}=\frac{3}{8}(1-w)p q^{-\tfrac{1+w}{1-w}}\,.
\end{equation}
 As the singularity is approached $V\rightarrow 0$ and $\theta\rightarrow -\infty$ or $+\infty$.
%%%%%%%%%%%%%%%%%%%%%%%%%%%%%%%%%%%%%%%%%%%%%%%%
%%%%%%%%%%%%%%%%%%%%%%%%%%%%%%%%%%%%%%%%%%%%%%%
\section{Quantization of the half-plane}
Because we are going to use affine transformations of physical quantities, we should keep control of the physical dimensions. In view of this, we introduce the  parameter $\sigma= a_P^{3(1-w)/4}$ which has the physical dimension of $q$. The dimensionless scale-momentum half-plane is then defined as $\Pi_+ = \{(\mq,\mp)\, |\, \mp\in \R\, , \, \mq> 0\}$, where
\begin{equation}
\label{scaledef}
\mq:= \frac{q}{\sigma}\, \quad \mp:= \sigma\frac{p}{a_P}
\end{equation}
Equipped with the multiplication 
\begin{equation}
(\mq,\mp)\cdot(\mq_0,\mp_0)=\left(\mq \mq_0, \frac{\mp_0}{\mq}+\mp\right), ~\mq\in\R^{\ast}_+, ~\mp\in\R\,, 
\end{equation}
it is viewed as the  affine group Aff$_{+}(\R)$ of the real line (see Appendix \ref{B} for more details).  This group possesses
two non-equivalent unitary irreducible representations (UIR) $U_{\pm}$,  besides the trivial one. Both are square integrable and this property is fundamental for the \textit{continuous wavelet analysis} \cite{gelnai,aslaklauder,-grosmor,-gros1,-gros2,G2000}. 
Equivalent realizations of one of them, say $U_+\equiv U$,  are carried on by  Hilbert spaces $\mathcal{H}_{\alpha} = L^2(\R^{\ast}_+, \ud x/x^{\alpha +1})$ where $\alpha$ is an arbitrary real number. Nonetheless, detailed calculations prove that these multiple possibilities do not introduce noticeable differences. Therefore we choose in the sequel the standard case $\alpha= -1$ and denote $\mathcal{H} =\mathcal{H}_{-1}=L^2(\R^{\ast}_+, \ud x)$. The UIR of Aff$_{+}(\R)$, expressed in terms of the physical phase space variables,  acts on $\mathcal{H}$ as 
\begin{equation}
\label{uiraff}
U(q,p) \psi(x) = e^{i\mp x } \frac{1}{\sqrt{\mq}} \psi(x/\mq)\,. 
\end{equation} 

Given a \underline{normalized} vector $\psi_0 \in \mathcal{H}$, a continuous family of unit vectors  are defined as 
\begin{equation}
\label{wavelet}
|q,p\rg = U(q,p)|\psi_0 \rg\, , \quad \lg x | q,p\rg = e^{i\mp x}\frac{1}{\sqrt{\mq}} \psi_0(x/\mq)\,. 
\end{equation}
where orthonormal basis $|x\rg$  in a distributional  sense obeys $\lg x|y\rg = \delta (x-y)$. The transported  $ \psi_0$ is named in signal analysis  (mother) wavelet or fiducial vector. 
For the sake of simplicity we assume in the sequel $\psi_0(x)$ is real-valued (for a discussion of complex-valued fiducial vectors, see Appendix \ref{C}).\\
Let us define the constants $c_\alpha$ (for real $\alpha$) as
\begin{equation}
\label{cdef}
 c_{\alpha} :=  \int_0^{\infty} \psi_0 (x)^2\, \frac{\ud x}{x^{2+\alpha}},
\end{equation}
the resolution of the unity is straightforward:
\begin{equation}
\label{resuniwave}
\int_{\Pi_+} \frac{\ud q \ud p}{2\pi \ap c_{-1}} | q,p\rg\lg q,p| = \lu,  
\end{equation}
provided that $c_{-1} < \infty$, which means that $\psi \in L^2(\R^{\ast}_+, \ud x/x)$, or, equivalently, that its Fourier transform is of null average, a classical requirement of continuous wavelet analysis.
Due to this crucial property, the family (\ref{wavelet}) is called  continuous wavelet basis (in signal analysis) or coherent state family for the affine group within a more quantum oriented context. 
Thanks to (\ref{resuniwave}), the CS quantization of classical  functions $f(q,p)$ can be implemented through
\begin{equation}
\label{quantfaff}
f\mapsto A_f = \int_{\Pi_+} \frac{\ud q \ud p}{2\pi \ap c_{-1}}\, f(q,p) \, | q,p\rg\lg q,p|\, \, . 
\end{equation}
Indeed, when properly defined, this map is linear, to $f=1$ there corresponds the identity operator and to real semi-bounded $f$ there corresponds a symmetric operator with self-adjoint extension(s), i.e. the three basic requirements of any quantization procedure. 

By construction, the map \eqref{quantfaff} is \emph{covariant} with respect to the unitary affine action $U$:
\begin{equation}
\label{covaff}
U(q_0,p_0) A_f U^{\dag}(q_0,p_0) = A_{\mathfrak{U}(q_0,p_0)f}\, , \quad \left(\mathfrak{U}(q_0,p_0)f\right)(q,p)=
f\left((q_0,p_0)^{-1}(q,p)\right)\, , 
\end{equation}
$\mathfrak{U}$ being the left regular representation of the affine group.
%%%%%%%%%%%%%%%%%%%%%%%%%%%%%%%%%%%%%%%%%
%%%%%%%%%%%%%%%%%%%%%%%%%%%%%%%%%%%%%%%%%%
\section{Physical operators}

\subsection{Functions of $q$}

The quantization of coordinate functions reads as 
\begin{equation}
\label{quantq}
q \mapsto A_q = \frac{c_0}{c_{-1}} Q\, , \quad Q\phi(x) := \sigma x\phi(x)\, , 
\end{equation}
provided that $c_0<\infty$. 
This operator is self-adjoint.\\
More generally for $\beta \in \R$
\begin{equation}
q^\beta \mapsto A_{q^\beta} = \frac{c_{\beta-1}}{c_{-1}}Q^\beta \, , 
\end{equation}
provided that $c_{\beta-1}<\infty$.\\
In particular the quantization of the volume $V= q^{\tfrac{2}{1-w}}$ reads as
\begin{equation}
A_V = \frac{c_{\frac{1+w}{1-w}}}{c_{ -1}} Q^{\tfrac{2}{1-w}}
\end{equation}

\subsection{Functions of $p$}

The quantization of momentum reads as
\begin{equation}
\label{quantp}
p \mapsto A_p = P\, ,  \quad P\phi(x) := -i \frac{\ap}{\sigma} \phi'(x)
\end{equation}
This operator is symmetric when defined on rapidly decreasing functions  with support in $\R^{\ast}_+$, but one can show that there is no self-adjoint extension (the deficiency indices are (1,0) \cite{reedsimon}).

One notices that the affine quantization yields canonical commutation rule $\left\lbrack A_q,A_p \right\rbrack= i \ap c_0/c_{-1}$.

For the ``\emph{free}" Hamiltonian $H_0=A_{p^2}$ we have:
\begin{equation}\label{defK}
H_0 = P^2 + \ap^2 \frac{K}{Q^2}\, ,  \quad \textrm{with} \quad K=K(\psi_0):=\int_0^{\infty}\frac{u \ud u}{c_{-1}}  (\psi_0'(u))^2 
\end{equation}
We notice that the quantization procedure always yields an additional term. This term depends only on the fiducial vector and its importance will be explained below.
\subsection{The quantized Hamiltonian} The classical Hamiltonian $h_T$ of \eqref{frweqs} reads as
\begin{equation}
h_T(q,p)= \alpha(w) p^2 + 6 \tilde{k} q^{\mu(w)} \, , \ \textrm{with} \ \tilde{k}=\left( \smallint \ud\omega \right) ^{2/3} k, \, k=-1,0,1\, ,
\end{equation}
therefore the quantum Hamiltonian $H=A_{h_T}$ reads as
\begin{equation}
\label{Htotal}
H=\alpha(w) P^2 +  \ap^2 \alpha(w) \frac{K}{Q^2} + 6 \tilde{k}  \frac{c_{\mu(w)-1}}{c_{-1}} Q^{\mu(w)}\, .
\end{equation}
Let us notice that $K>0$, whatever the choice of $\psi$.  Therefore the singularity $x=0$ cannot be reached: this dressing of the classical singularity is an outcome of the followed CS quantization scheme (through $K$). Furthermore our procedure yields a renormalization of the coupling constant of the potential.\\
If we assume a closed universe  with a radiation content then $w=1/3$, $k=+1$ and $\tilde{k}=(\int \ud\omega)^{2/3}$, then variables $q$ and $p$ get both a length dimension (Eq.(\ref{dimqp})) and  we obtain the special Hamiltonian $H_{cr}$
\begin{equation}
\label{Hspecial}
H_{cr}=\frac{1}{24} P^2 +  \frac{ \ap^2 K}{24} \frac{1}{Q^2} + 6 \tilde k\frac{c_{1}}{c_{-1}} Q^2\, .
\end{equation}
This Hamiltonian is an ordinary differential Sturm-Liouville operator, singular at the end point $x=0$. The functional properties depend on the value of $K$, as follows from the analysis of Gesztesy {\it et al} \cite{Gesztesy} (see also \cite{reedsimon}). In particular, $K=3/4$ is the critical value, while one would naively expect $K=0$, i.e., the infinite barrier, to play the role. Using the standard approach and terminology of  \cite{reedsimon}, the potential term in $H_{cr}$ is in the limit point case at the end point $x = 0$ if $K \ge 3/4$ and in the limit circle case at $x=0$ if $0 \le K < 3/4$. The potential is in the limit point case at infinity. It follows that $H_{cr}$ (defined on the domain of smooth compactly supported functions) is essentially self-adjoint in the former case. In the latter range of $K$, the deficiency indices of $H_{cr}$ are $(1, 1)$ and therefore more self-adjoint extensions exist; see \cite{Gesztesy} for a detailed analysis. In this paper, we choose the fiducial vector $\psi_0$ in such a way that $K \ge 3/4$ in order to comply with essential self-adjointness of the Hamiltonian. This ensures unambiguous time evolution and singularity resolution at the quantum level.

\subsection{The quantum expansion rate $\Theta$} We have
\begin{equation}
\theta =\frac{3}{8}(1-w)p q^{-\frac{1+w}{1-w}} \mapsto A_\theta = \Theta = \frac{3 c_{-\frac{2}{1-w}}}{8\, c_{-1}} (1-w)\frac{1}{Q^{\frac{1+w}{2(1-w)}}} P \frac{1}{Q^{\frac{1+w}{2(1-w)}}}.
\end{equation}
This operator is symmetric when defined on a suitable domain, but it is not self-adjoint (or does not possess a self-adjoint extension), from the reasoning previously developed for $P$.
%%%%%%%%%%%%%%%%%%%%%%%%%%%%%%%%%%%%%%%%%%%%%%
%%%%%%%%%%%%%%%%%%%%%%%%%%%%%%%%%%%%%%%%%%%%%
\section{Explicit formulae in a particular case} In the previous formula the wavelet $\psi_0$ is a free parameter of our CS quantization. In order to obtain explicit expressions, we decide to fix $\psi_0$  as the following unit vector in $\mathcal{H}$:
\begin{equation}
\psi_0^{\nu,\xi}(x)= \frac{1}{\sqrt{2 x \, K_0(\nu)}}e^{-\frac{ \nu}{4} \left(\xi x +\frac{1}{\xi x}\right)}, \quad \textrm{with} \,\, \nu>0 \, \  \textrm{and} \,\ \xi >0.
\end{equation}
We notice that $\psi_0^{\nu,\xi}(x)$ falls off with all its derivatives at the origin and at the infinity. To calculate the normalization constant involved in the previous definition of $\psi_0^{\nu,\xi}$ (and in order to obtain other useful integrals) we take benefit of the formula \cite{Gradshteyn}
\begin{equation}
\forall a, b, c \in \mathbb{C}, \, \Re(b) >0, \Re(c) >0, \quad \int_0^\infty x^{a-1} e^{-c x-b/x} \ud x = 2 \left( \frac{b}{c} \right)^{a/2} K_{-a}(2 \sqrt{b c})\, . 
\end{equation}
Here, $K_a$ is a modified Bessel function \cite{magnus66}. We recall that its asymptotic behavior at large argument $\nu$ is $K_a(\nu) \sim e^{-\nu}\sqrt{\pi/(2\nu)}$, whereas at small $\nu\ll \sqrt{a+1}$, $K_a(\nu) \sim (1/2)\Gamma(a)(2/\nu)^a$ for $a>0$ and $K_0(\nu) \sim -\ln(\nu/2) -\gamma$.\\
The coefficients $c_{\alpha}$ defined in Eq.\eqref{cdef} reads as
\begin{equation}
c_\alpha = \frac{\xi^{\alpha+2}}{K_0(\nu)} K_{\alpha+2}( \nu).
\end{equation}
The coefficient $\xi$ is fixed in the sequel as being
\begin{equation}
\xi = \frac{K_1(\nu)}{K_2(\nu)}.
\end{equation}
The interest of this choice will become apparent with the study of the operator $A_q$. Also note  that it implies that, in view of the study of the semi-classical regime,  $a_P$   and $\nu$ remain the only free parameters.\\
The coefficient $c_{-1}$ involved into the resolution of unity is
\begin{equation}
c_{-1}=\frac{K_1(\nu)^2}{K_0(\nu)K_2(\nu)}.
\end{equation}
\subsection{The physical operators}
The quantization of the variable $q$ reads as
\begin{equation}
A_q = \frac{c_0}{c_{-1}} Q = Q \, , 
\end{equation}
due to of the choice of $\xi$.
The quantization of the  potential $q^\beta$  leads to
\begin{equation}
\label{quantqbeta}
A_{q^\beta}= \frac{c_{\beta-1}}{c_{-1}} Q^\beta = \frac{K_1(\nu)^{\beta-1} K_{\beta+1}(\nu)}{K_2(\nu)^\beta} Q^\beta.
\end{equation}
This formula can be applied in particular to the classical volume $V=q^{2/(1-w)}$.\\
The quantization of the momentum $P=A_p$ is as in (\ref{quantp})
while the ``\emph{free}" Hamiltonian $H_0=A_{p^2}$ becomes
\begin{equation}
H_0 = P^2+a_P^2 \frac{K(\nu)}{Q^2}\, , \quad  K(\nu):=\frac{1}{4}\left(1+\nu \frac{K_0(\nu)}{K_1(\nu)}\right).
\end{equation}
The Hamiltonian $H$ of Eq.\eqref{Htotal} reads as
\begin{equation}
\label{Hwithgammanu}
H = \alpha(w)  P^2+ a_P^2\alpha(w)\frac{K(\nu) }{Q^2} + 6 \tilde{k} \frac{K_1(\nu)^{\mu(w)-1} K_{\mu(w)+1}(\nu)}{K_2(\nu)^{\mu(w)}} Q^{\mu(w)} \,.
\end{equation}
The last quantized observable is the expansion rate $\Theta=A_\theta$ that reads as
\begin{equation}
\Theta =  \frac{3(1-w)}{8}K_{-\frac{2 w}{1-w}}(\nu) \frac{K_2(\nu)^{\tfrac{1+w}{1-w}}}{K_1(\nu)^{\tfrac{2}{1-w}}} \frac{1}{Q^{\tfrac{1+w}{2(1-w)}}} P \frac{1}{Q^{\tfrac{1+w}{2(1-w)}}} 
\end{equation}
Finally, as was already pointed out above (Eq. \eqref{Hspecial}), the special interesting case of Hamiltonian $H$  is obtained when we assume a closed universe with a radiation content ($k=+1$, $w=1/3$ and $\tilde{k}=(\int \ud\omega)^{2/3}$), and yields the Hamiltonian
\begin{equation}
\label{Hfinal}
H_{cr} = \frac{1}{24} P^2+ \frac{a_P^2}{24}\frac{K(\nu)}{Q^2}+ 6 \tilde{k}\frac{K_1(\nu)K_3(\nu)}{K_2(\nu)^2} Q^2.
\end{equation}

\subsection{Analysis of $H_{cr}$}

\subsubsection{Semi-classical point of view}
If we perform a semi-classical analysis of the dynamics due to $H_{cr}$ obtained in Eq.\eqref{Hfinal}, we see that the supplementary repulsive potential generated by the quantization leads to a displacement of the equilibrium point of the potential. While the harmonic potential alone possesses an equilibrium point located at the singularity $q=0$, the harmonic potential corrected with the supplementary repulsive term exhibits a different equilibrium point which is located at 
\begin{equation}
q_e^4=\frac{a_P^2}{144} \frac{K_2(\nu)^2}{K_1(\nu)K_3(\nu)} \, K(\nu)\,.
\end{equation}
We recover $q_e = 0$ when $a_P = 0$, but for $a_P \ne 0$ we find that the smallest value $q_e \simeq \sqrt{a_P}/5.8$ is obtained for $\nu \to 0$, while $q_e \to \infty$ when $\nu \to \infty$. Moreover the renormalized coupling constant of the potential in Eq.\eqref{Hfinal} converges to classical counterpart when $\nu \to \infty$. Therefore the free parameter $\nu$ and the standard $a_P$ can be used to specify  the renormalized coefficient of the potential and the position of the new equilibrium point. \\
Hence we observe that the semi-classical dynamical behavior of the system is an oscillation around this point $q_e$ and the ``bare (or true) classical singularity $q=0$'' is never reached.

\subsubsection{Eigenstates and evolution operator}
With the quantities $\ell$, $\omega$ and $\lambda$ defined as
\begin{equation}
\ell = \frac{1}{2} \left( \sqrt{2+ \nu \frac{K_0(\nu)}{K_1(\nu)}}-1\right)\,, \quad \omega=\frac{\sqrt{\tilde k K_1(\nu) K_3(\nu)}}{K_2(\nu)} \, , \, \textrm{and} \,\, \lambda= 6 \omega\,,
\end{equation}
the eigenvalues $E_n$, $n=0,1, \dots$ of $H_{cr}$ read as
\begin{equation}
\label{eigenvalues}
E_n = a_P \omega (2n+\ell+3/2)\, .
\end{equation}
The corresponding normalized eigenvectors $\psi_n$ are
\begin{equation}
\label{eigenstates}
\psi_n(x) = N_n \, x^{\ell+1} L_n^{\ell+1/2}(2 \lambda x^2) e^{-\lambda x^2}\,.
\end{equation}
 The functions $L_n^{\ell+1/2}$ are the associated Laguerre polynomials \cite{magnus66} and the normalization factor $N_n$ is given by
\begin{equation}
N_n = \left( \frac{2 (2 \lambda)^{\ell+3/2} \, n!}{\Gamma(n+\ell+3/2)} \right)^{1/2}\,.
\end{equation}
Introducing a dimensionless evolution parameter $\tau$, related to time $t$ through some scaling for instance,  the evolution operator  $U(\tau)=e^{-i H_{cr} \tau/a_P}$ is periodic with period $\pi/\omega$. It  is given in terms of \emph{matrix elements} $\bra{x} U(\tau) \ket{y}$ by:
\begin{equation}
\label{evolop}
\bra{x} U(\tau) \ket{y} = 2 \lambda  \frac{\sqrt{x y}}{|\sin(\omega \tau)|}  e^{i \lambda (x^2+y^2) \cot(\omega \tau)} J_{\ell+1/2} \left( \frac{\lambda x y}{|\sin(\omega \tau)|}\right)\, ,
\end{equation}
where $J_{\nu}$ is a Bessel function. This expression is derived from  series  involving Laguerre polynomials \cite{Gradshteyn}. The r.h.s. of Eq. \eqref{evolop} is singular for $\sin(\omega \tau)=0$,  $\bra{x} U(\tau) \ket{y}$ being in that case the distribution $\delta(x-y)$, corresponding to $U(\tau)=1$.

\subsection{Lower symbols} The expectation values  $\lg q,p | A| q,p \rg$ of quantum operators allow to map the quantum world to the classical one and they are called ``lower symbols" \cite{lieb}  or ``covariant symbols'' \cite{berezin}. Therefore we can examine the \emph{semi-classical map} $f(q,p) \mapsto  \check{f}(q,p)=\lg q,p | A_f| q,p \rg$ that exhibits the corrections (regularizations) induced from our quantization procedure.  The map $f \mapsto \check{f}$ reads as the integral transform
\begin{equation}
\check{f}(q,p) = \int_{\Pi_+} \frac{\ud p' \ud q'}{2\pi a_P c_{-1}} |\lg q,p|q',p' \rg |^2 f(q',p')\, .
\end{equation}
It is the average value of the function $f(q,p)$ with respect to the probability distribution $(q',p') \mapsto \frac{1}{2\pi a_P c_{-1}} |\lg q,p|q',p' \rg |^2$. Viewed as a kernel, the latter is expected to play a regularizing role under the form of a generalized convolution.
Due to the resolution of the unity (\ref{resuniwave}), the scalar product $\lg q,p|q',p' \rg$ is a reproducing kernel for a Hilbert subspace of   $L^2(\Pi_+, \ud q\,\ud p)$. It is given by
\begin{equation}
\lg q,p|q',p' \rg = \frac{1}{K_0(\nu)} K_0\left( \nu \frac{q+q'}{2 \sqrt{qq'}} \sqrt{1+\frac{4i q q'(p'-p)}{a_P \nu \xi (q+q')}} \right)\, , \, \textrm{with} \ \xi=\frac{K_1(\nu)}{K_2(\nu)}\, .
\end{equation}
For the powers of $q$ we have
\begin{equation}
\label{lowerqbeta}
\lg q,p|A_{q^\beta}|q,p \rg = \frac{K_\beta(\nu) K_{\beta+1}(\nu)}{K_0(\nu)K_1(\nu)} q^\beta\, ,
\end{equation}
in particular 
\begin{equation}
\lg q,p|A_q|q,p \rg  = \lg q,p|Q|q,p \rg =\frac{K_2(\nu)}{K_0(\nu)} q\,.
\end{equation}
We recover $\lg q,p|A_q|q,p \rg \simeq q$ if $\nu \to \infty$.
Otherwise we obtain for $P=A_p$
\begin{equation}
\lg q,p|A_p|q,p \rg =\lg q,p|P|q,p \rg =p.
\end{equation}
The expectation value $ \lg q,p|P^2|q,p \rg $ reads as
\begin{equation}
\label{lowerpsquare}
\lg q,p|P^2|q,p \rg =p^2 + a_P^2 \frac{K_1(\nu)^2 K(\nu)}{ K_0(\nu) K_2(\nu)} \frac{1}{q^2}\,.
\end{equation}
We notice the supplementary term in $1/q^2$. Using the previous equations \eqref{quantqbeta}, \eqref{lowerqbeta} and \eqref{lowerpsquare} we obtain
\begin{equation}
\lg q,p|A_{p^2}|q,p \rg = p^2+ 2a_P^2 \frac{K_1(\nu)^2K(\nu)}{ K_0(\nu) K_2(\nu)} \frac{1}{q^2}\,.
\end{equation}
We notice that the $1/q^2$ coefficient in $\lg q,p|A_{p^2}|q,p \rg$ and in $\lg q,p|P^2|q,p \rg$ differ from a factor $1/2$.

The coupling constant of the supplementary term in $q^{-2}$ vanishes as $a_P \to 0$ or $\nu \to 0$, it becomes infinite as $\nu \to \infty$ (for a fixed value of $a_P$). Finally we obtain for the classical Hamiltonian $h_T$ of Eq.\eqref{frweqs}
\begin{equation}
\label{lowerH}
\check{h}_T(q,p) =\alpha(w) p^2 +2a_P^2\alpha(w)\frac{ K_1(\nu)^2 K(\nu)}{ K_0(\nu) K_2(\nu)}\frac{1}{q^2} + 6 \tilde{k} \frac{K_{\mu(w)}(\nu) K_{\mu(w)+1}(\nu)}{K_0(\nu)K_1(\nu)} q^{\mu(w)}.
\end{equation}
We recover $\check{h}_T \to h_T$ (and $\check{q^\beta} \to q^\beta$, $\check{p^2} \to p^2$) if we assume first $a_P \to 0$ and then $\nu \to \infty$ (the limits are not commuting). If we want to keep two independent parameters and to have independent limits, we need to ``renormalize" $a_P$, including in its definition some of the $K$-Bessel functions involved in the coupling coefficient of $q^{-2}$ as for example
\begin{equation}
a_P = \frac{\tilde a_P}{2} \sqrt{\frac{K_0(\nu)K_2(\nu)}{K_1(\nu)^2 K(\nu)}},
\end{equation}
where $\tilde a_P$ would be the new renormalized Planck area. Another solution (to avoid the problem of non commuting limits) is to assume that $\nu$ is in fact a function of $a_P$. The expression $\nu(a_P)$ must be well chosen: we need to impose both $a_P^2 \nu(a_P) \to 0$ and $\nu(a_P) \to \infty$ as $a_P \to 0$ (a simple solution is $\nu(a_P)\propto 1/a_P$).

\subsection{Time evolution in phase space} For any normalized state $\phi \in \mathcal{H}$, the resolution of unity allows us to get its phase space representation
\begin{equation}
\label{distprobphsp}
\Phi(q,p)  =\frac{1}{\sqrt{2\pi a_P c_{-1}}} \lg q,p | \phi \rg 
\end{equation}
and the resulting   probability distribution on the phase space $\Pi_+$:
\begin{equation}
\label{eqn:phasespacerepre}
\Pi_+ \ni (q,p)\mapsto  \frac{1}{2\pi a_P c_{-1}} |\scalar{q,p}{\phi}|^2=\rho_{\phi}(q,p)\, .
\end{equation}

%%%%%%%%%%%%%%%%%%%%%%%%%%%%%%
\begin{figure}[!ht]
\begin{tabular}{cc}
\includegraphics[scale=0.37]{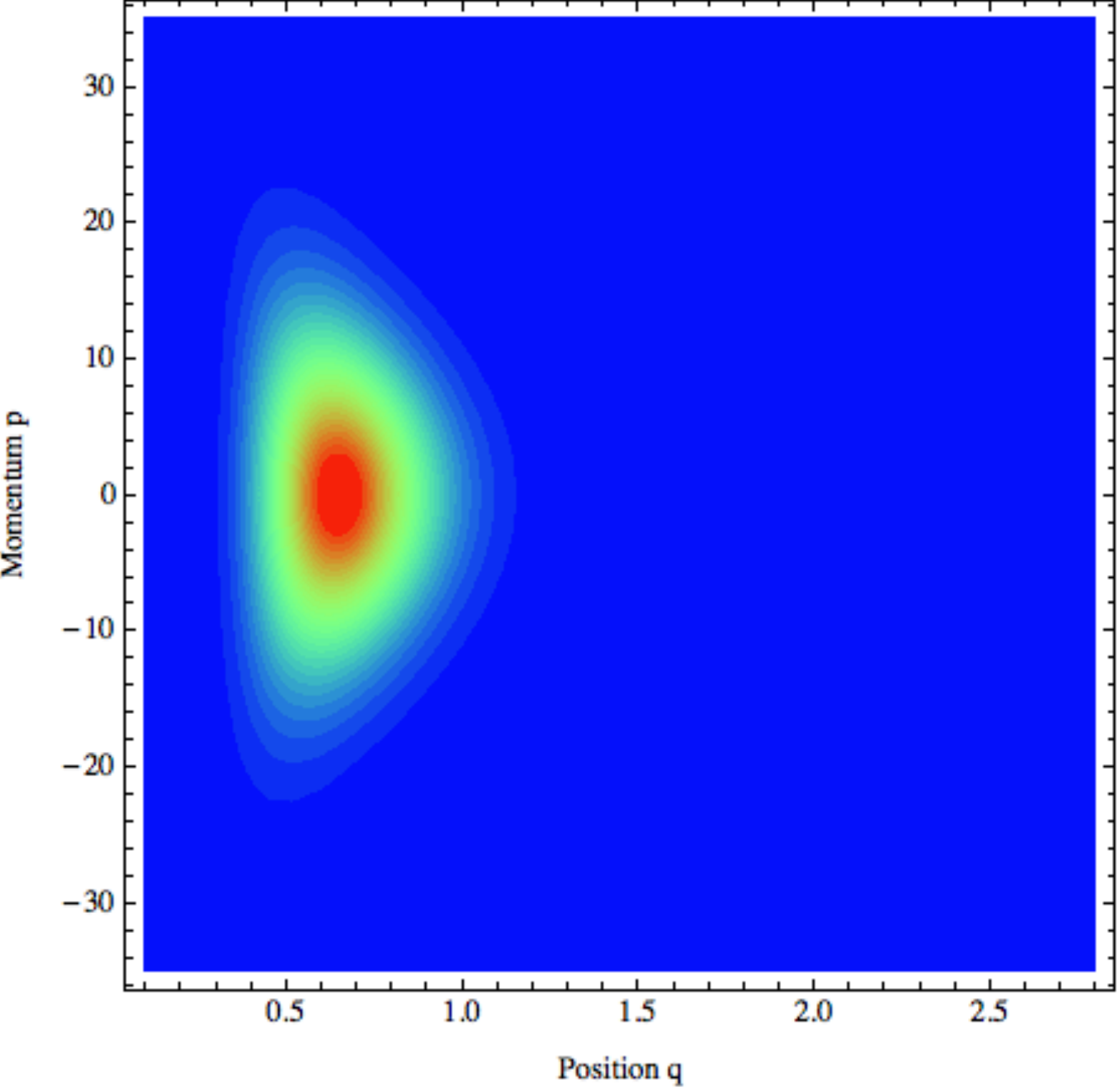} & 
\includegraphics[scale=0.37]{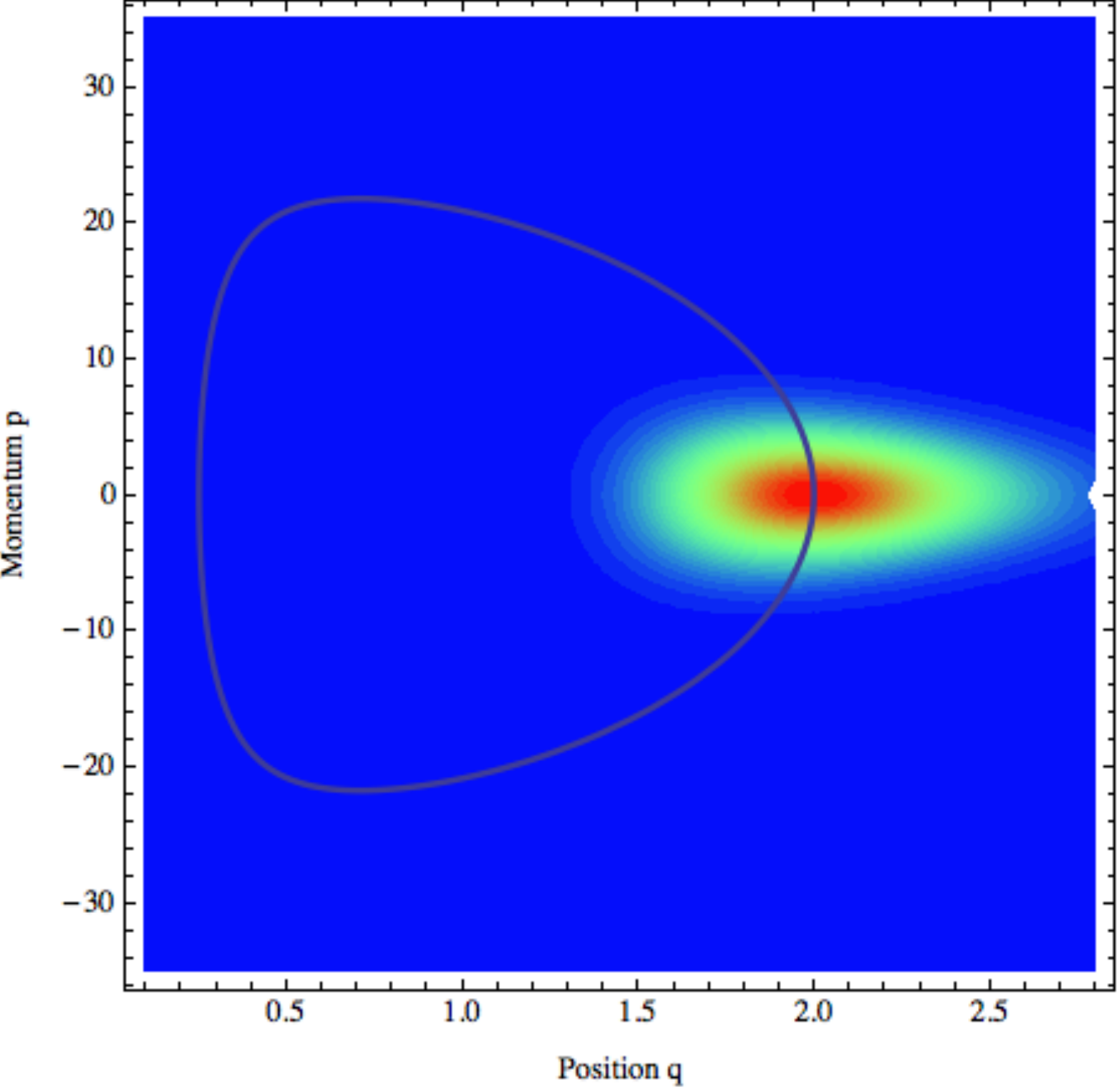}
\end{tabular}
\caption{Phase space distributions \eqref{eqn:phasespacerepre}: on the left for the eigenstate $\psi_0$ of $H_{cr}$, on the right  for the coherent state $| q_0,p_0 \rg$ with $q_0=2$, 
$p_0=0$. The parameters are fixed to the values $\nu=80$, $a_P=1$ and $\tilde{k}=1$. The thick curve (figure on the right) is the expected phase trajectory, deduced from the semi-classical hamiltonian in Eq. \eqref{eqn:semiclassenergy}. The ranges of variables $q$ and $p$ are respectively $[0.2,\, 2.8]$ and $[-35,\, +35]$. Increasing values of the function are encoded by the colors from blue to red. } 
\label{figure1}
\end{figure}
%%%%%%%%%%%%%%%%%%%%%%%%%%%%%%%

In Figure \ref{figure1} two phase-space distributions $\rho_{\phi}(q,p)$ are shown: for the ground state $\psi_0$ of  $H_{cr}$ and for the coherent state $|q_0,p_0 \rg$ with $(q_0,p_0)=(2,0)$. In both cases the quantization parameters are  chosen as $\nu=80$, $a_P=1$ and $\tilde{k}=1$. 

Let us now examine the time behavior $\tau \mapsto \rho_{\phi(\tau)}(q,p)$ for a state $\phi(\tau)$ evolving under the action of the Hamiltonian $H_{cr}$ of Eq.\eqref{Hfinal}:
\begin{align}
\ket{\phi(\tau)}&=e^{-i {H_{cr}} \tau/a_P} \ket{\phi} \nonumber \\
&=\sum_{n=0}^\infty e^{-i E_n \tau/ a_P} \scalar{\psi_n}{\phi} \ket{\psi_n}\, , 
\end{align}
where $\ket{\psi_n}$ is  eigenstate of $H_{cr}$ with eigenvalue $E_n$  given in Eqs. \eqref{eigenvalues}, \eqref{eigenstates}.

With  $\ket{\phi}=\ket{q_0,p_0}$ as an initial state and using Eq.\eqref{lowerH},  we have  for a given $\nu$
\begin{equation}
\label{eqn:semiclassenergy}
\lg q_0,p_0,\tau\,|H_{cr}|q_0,p_0,\tau \rg =\frac{1}{24} p_0^2 + a_P^2\frac{ K_1(\nu)^2 K(\nu)}{12 K_0(\nu) K_2(\nu)}\frac{1}{ q_0^2}+6 \tilde{k} \frac{K_2(\nu) K_{3}(\nu)}{K_0(\nu)K_1(\nu)} q_0^2\, .
\end{equation}
Since for large values of $\nu$ the lower symbols of $Q=A_q$ and $P=A_p$ correspond to their classical original functions $q$ and $p$, one can expect that the time average of the probability law $\rho_{\, \ket{q_0,p_0,\tau}}(p,q)$, defined in \eqref{eqn:phasespacerepre}, corresponds to some fuzzy extension in phase space of the classical trajectory corresponding to the time-independent Hamiltonian in the r.h.s. of Eq.\eqref{eqn:semiclassenergy}. Similar dynamical issues  of CS quantization are  encountered in the case of P\"oschl-Teller potentials \cite{csbergeron10_12}.

%%%%%%%%%%%%%%%%%%%%%%%%%%%%%%
\begin{figure}[!ht]
\includegraphics[scale=0.4]{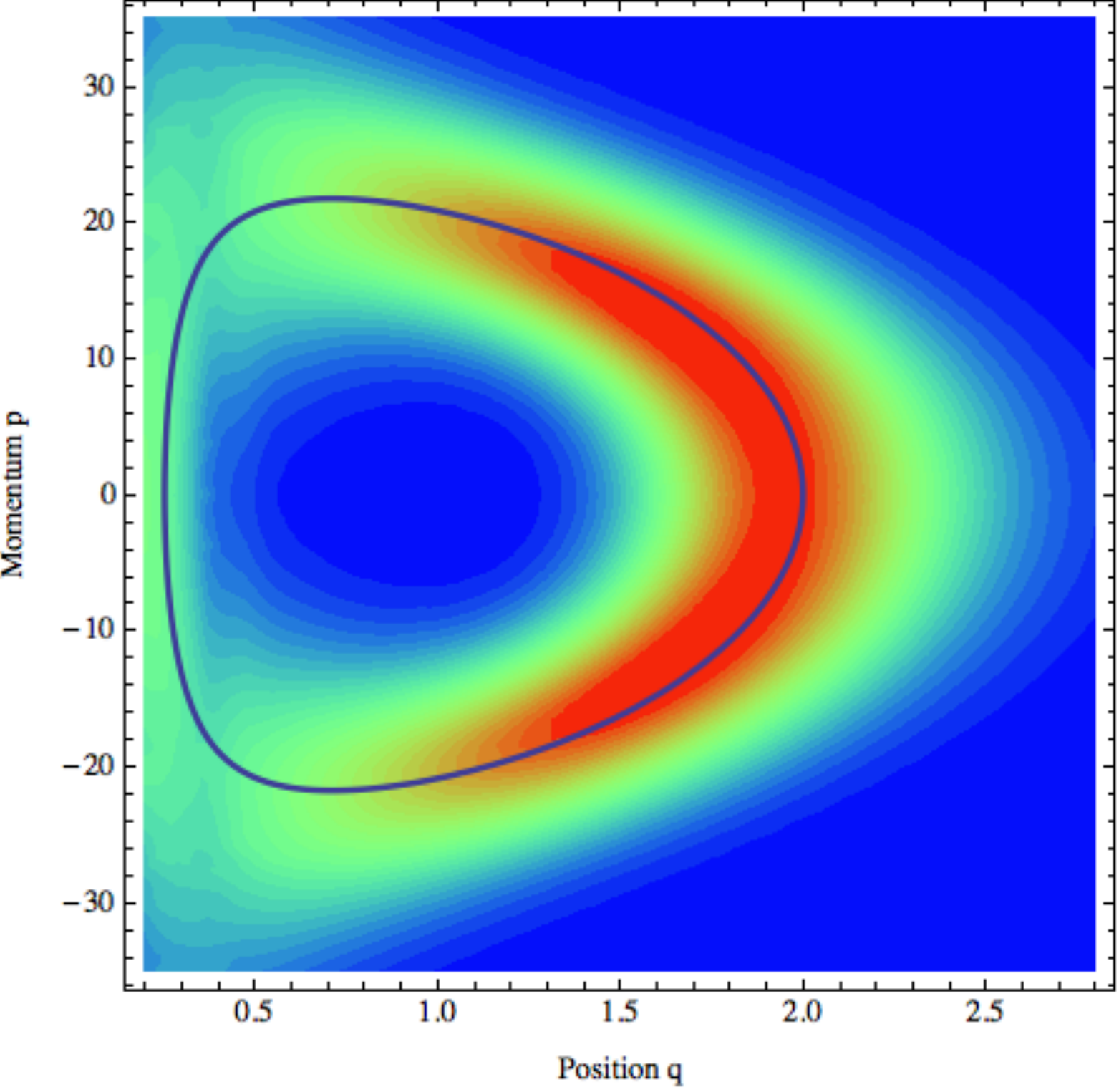}
\caption{Time average phase space distribution of Eq.\eqref{rhobar} for  $\ket{q_0, p_0}(\tau)$ evolving following the Hamiltonian $H_{cr}$. The values of parameters and the thick curve are those of Fig.\ref{figure1}. The ranges in $q$ and $p$ are respectively $[0.2,\, 2.8]$ and $[-35,\, +35]$. The domain $0 \le q < 0.2$ is not represented because of numerical instabilities. Increasing values of the function are encoded by the colors from blue to red. } 
\label{figure2}
\end{figure}
%%%%%%%%%%%%%%%%%%%%%%%%%%%%%%%

\begin{figure}[!ht]
\begin{tabular}{ccc}
\includegraphics[scale=0.37]{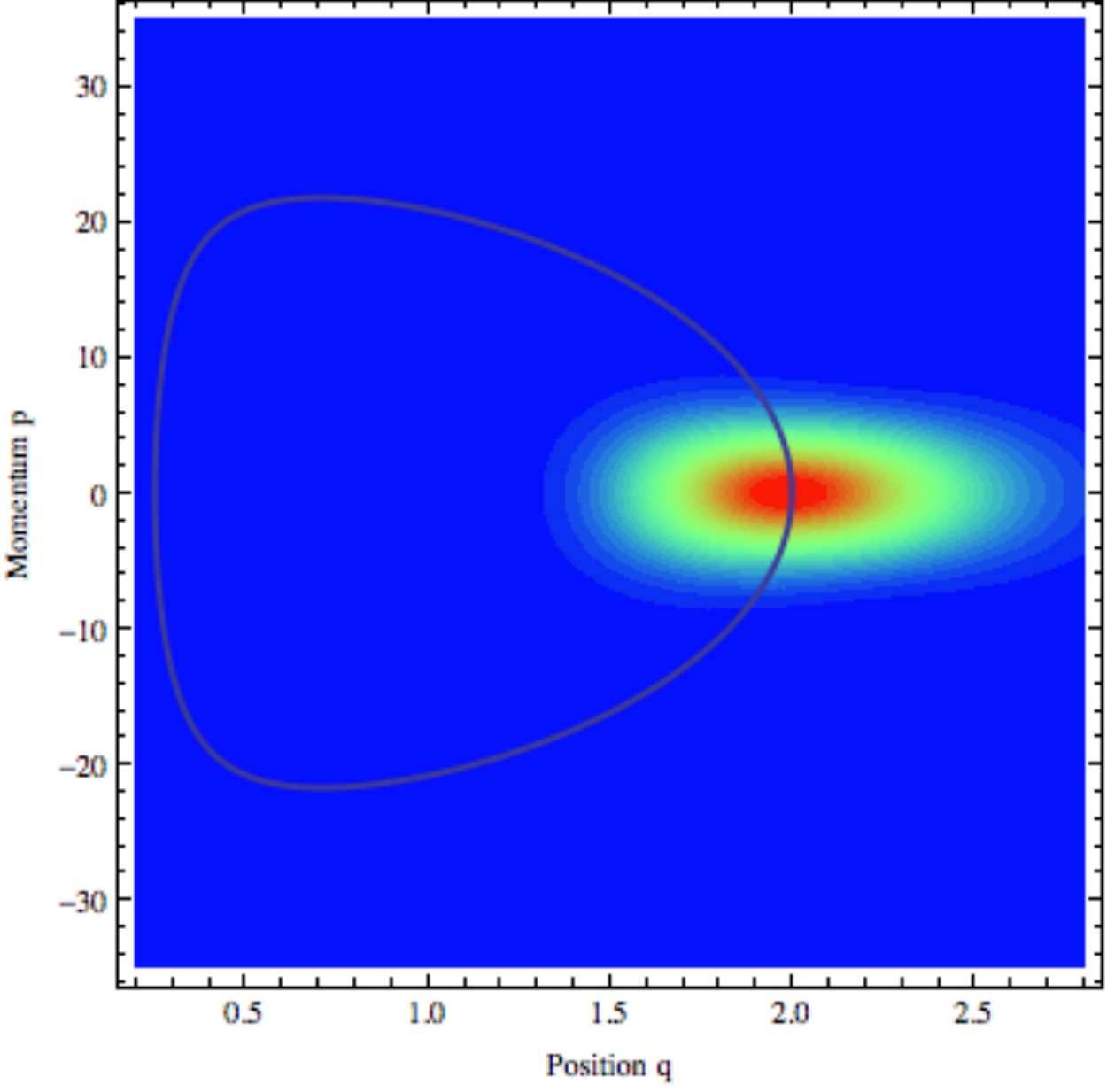} & 
\includegraphics[scale=0.37]{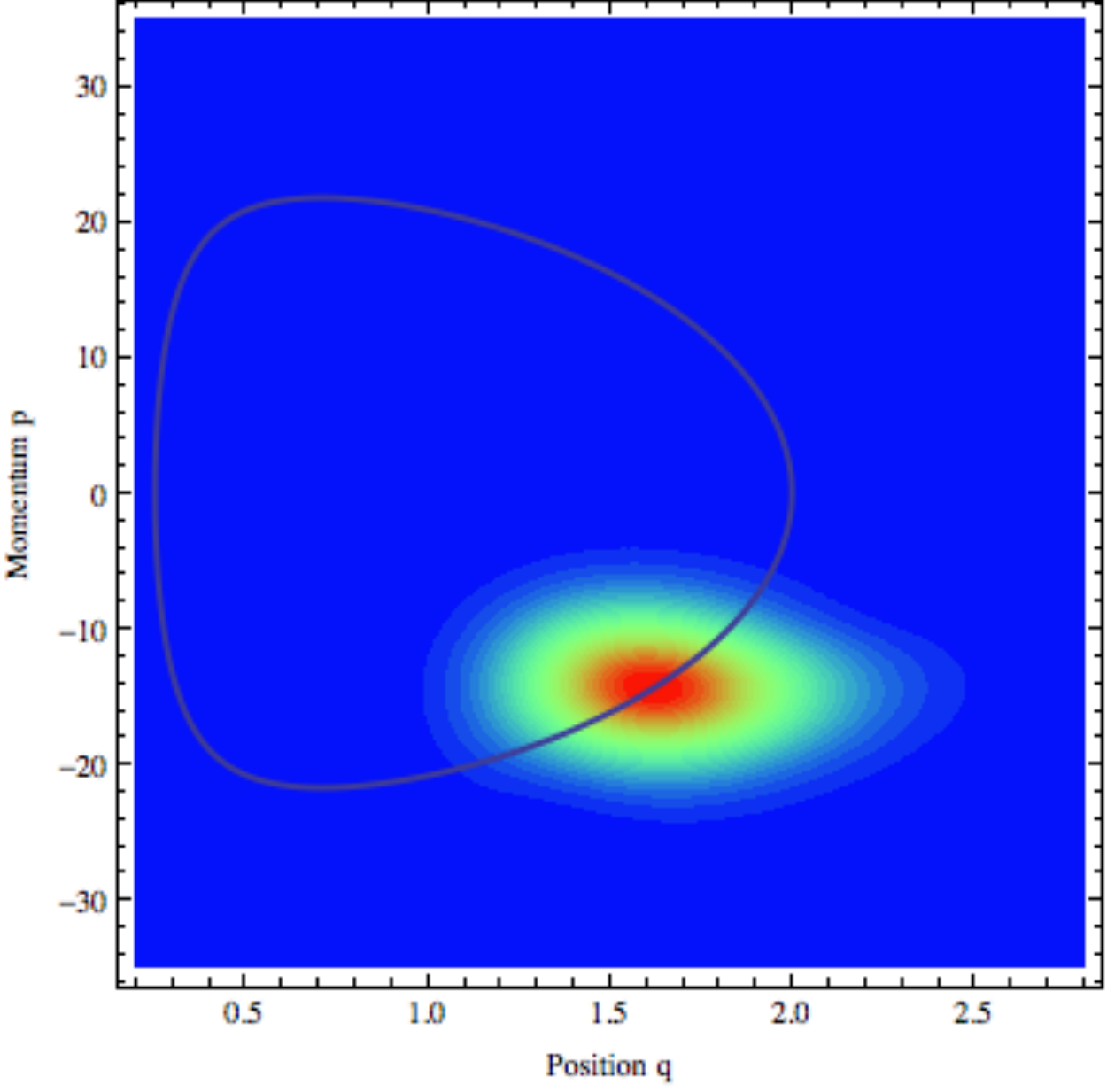} &
\includegraphics[scale=0.37]{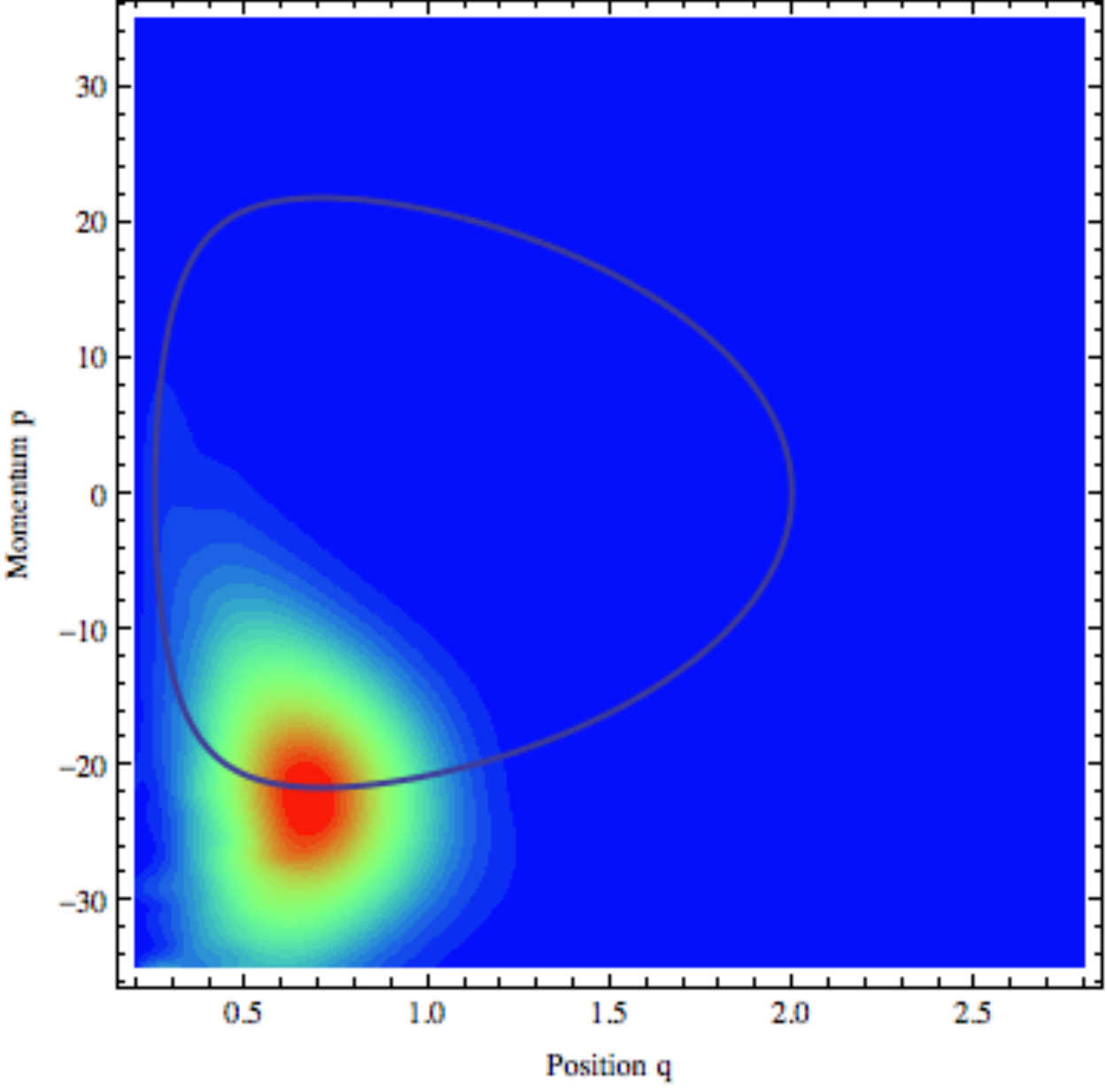} \\
\includegraphics[scale=0.37]{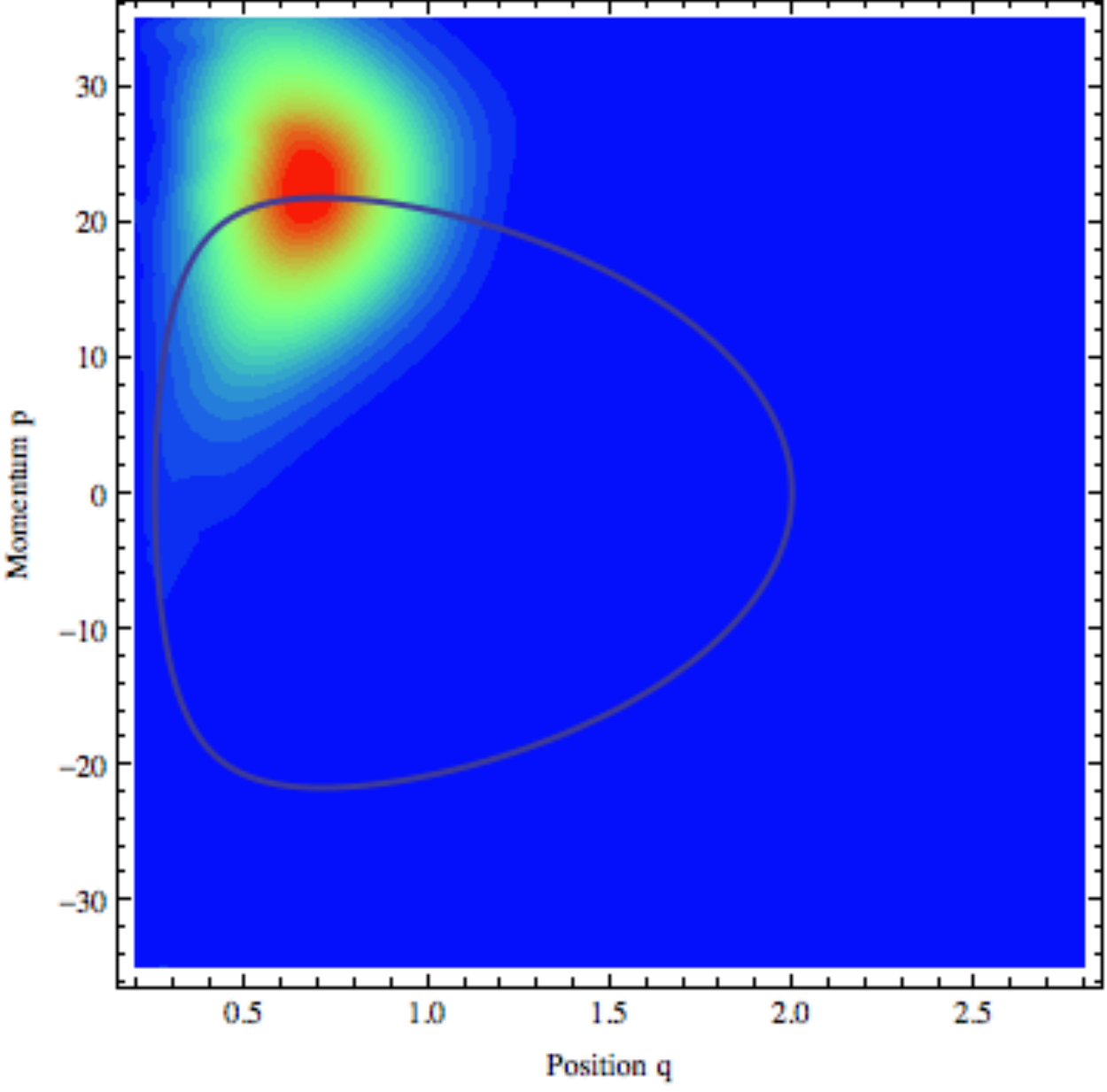} & 
\includegraphics[scale=0.37]{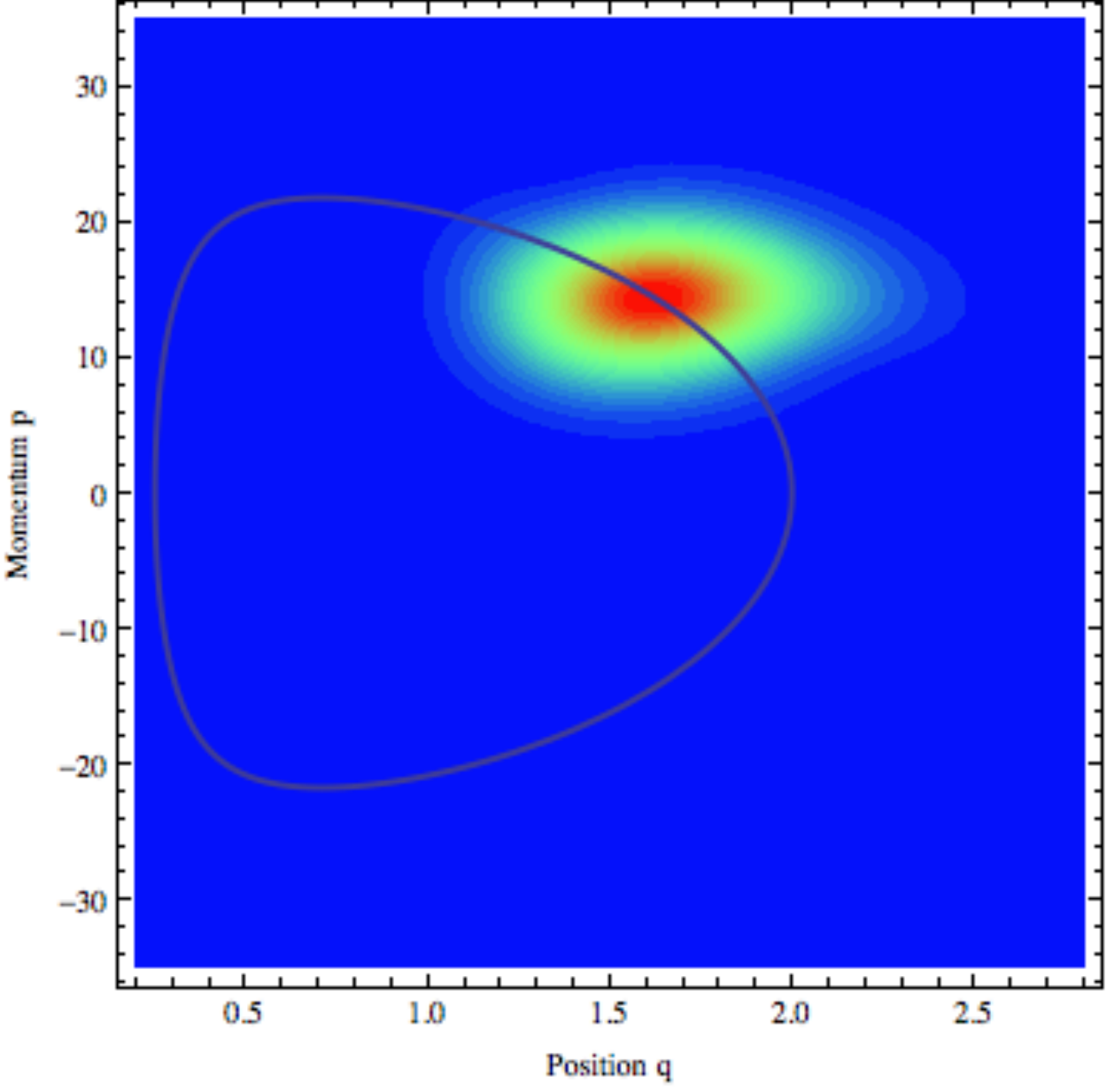} &
\includegraphics[scale=0.37]{Movie0.pdf} 
\end{tabular}
\caption{Phase space distributions $\rho_{\, \ket{q_0,p_0,\tau}}(p,q)$ at different times equally spaced (from top left to bottom right).  The values of parameters and the thick curve are those of Fig.\ref{figure1}. The ranges in $q$ and $p$ are respectively $[0.2,\, 2.8]$ and $[-35,\, +35]$. Increasing values of the function are encoded by the colors from blue to red.} 
\label{figure3}
\end{figure}

This key result is illustrated in Figure \ref{figure2} where we have represented the time average  distribution $\bar{\rho}$ defined as
\begin{equation}
\label{rhobar}
\bar{\rho}(q,p)=\lim_{T \to \infty} \frac{1}{T} \int_0^T \rho_{\ket{q_0,p_0,\tau}}(q,p) d\tau = \frac{1}{2\pi a_P c_{-1}} 
\sum_{n=0}^\infty |\scalar{q,p}{\psi_n}|^2 |\scalar{\psi_n}{q_0,p_0}|^2\, ,
\end{equation}
for the same values of the parameters as in Figure \ref{figure1}. 
The time average distribution $\bar{\rho}$ allows us to compare the quantum behavior with the classical trajectory, but the expression \eqref{rhobar} hides the details of the wave-packet dynamics, i.e. the bouncing of the wave-packet during its periodic motion. The figure \ref{figure3} represents this behavior.

%%%%%%%%%%%%%%%%%%%%%%%%%%%%%%
%%%%%%%%%%%%%%%%%%%%%%%%%%%%%%%

%%%%%%%%%%%%%%%%%%%%%%%%%%%%%%%%%%%%%%%%%%%%%%%
%%%%%%%%%%%%%%%%%%%%%%%%%%%%%%%%%%%%%%%%%%%%
\section{Discussion and Conclusion}

In this paper we employed integral quantization based on the affine coherent states to derive quantum models of homogeneous and isotropic universe. As we already noticed, the Weyl quantization brings unsatisfactory results, particularly when one considers essential self-adjointness of the quantum hamiltonian. This invites departures from the canonical prescription. Our approach, based on the group structure of the phase space, the affine group, provides a general quantization procedure for an arbitrary observable and preserves the basic commutation rule. In the obtained quantum model of the Friedman-Lemaitre universe the classical big bang singularity is replaced with a quantum big bounce resulting in a smooth and complete evolution. At the most general level, the removal of singularity is due to a unitary evolution: a gravitational collapse, represented by the Dirac delta peaked at $q=0$ can never be reached as this state does not belong to the Hilbert space. The novelty introduced by CS quantization is that the singularity resolution is accompanied by the occurrence of a repulsive potential. The potential's role is twofold: (i) on the quantum level it may lead to a non-ambiguous unitary evolution across the bounce, (ii) on the  semiclassical level it provides a mechanism for the universe to stop contraction, bounce and begin expansion, which is so much more natural than the hard bounce of the usual quantizations. The pre-factor in the potential term (which will be discussed  below) is the desired free parameter that provides a way to match our theory to observational constraints. In particular it may be used to set the energy scale at which quantum effects come into play. Let us emphasize that this inverse quadratic potential arises not only for {\it all} possible choices of the fiducial vector $\psi_0$ in our quantization scheme, but also with a quantization procedure issued from Weyl-Heisenberg coherent states.

\subsection{Finite universe}
In the case for which the universe is non-compact the Hamiltonian formulation is derived through the restriction to a finite patch of space. The inspection of the semiclassical Hamiltonian and the definition of the basic variables (\ref{bv}) and (\ref{nbv}) and the curvature constant $\tilde{k}$ shows the following: both the kinetic and curvature terms of the Hamiltonian depend on the size of the patch like $(\int \ud\omega)^{w+1}$ while the potential term behaves like $(\int \ud\omega)^{w-1}$. In other words, the classical dynamics is invariant with respect to the choice of the patch, whereas the  quantum dynamics is not. The repulsive potential breaks the invariance and so the {\it physical} content of the theory depends on the patch. Thus, one has to exclude non-compact universe from quantum modeling. We note that it does not imply the curvature of the universe: for $k=0$ we may consider e.g. torus topology, for $k=1$ the universe is necessarily compact, and for $k=-1$ there are infinitely many possible compact spatial sections \cite{Thur}.

\subsection{Planck era and the quantum phase} From the lower symbol of Hamiltonian (\ref{lowerH}) we deduce that the quantum effects become important once the repulsive potential gets comparable with the kinetic part, that is $p^2\approx (\beta^2(\nu)/q^2)$, where $$\beta^2(\nu)=2a_P^2 K_1(\nu)^2K(\nu)/(K_0(\nu)K_2(\nu))\,.$$ In other words, the region of phase space in which classical dynamics cannot be trusted is given by the inequality (with curvature neglected):
\begin{equation}\nonumber
|pq|\lesssim \beta(\nu)\, , 
\end{equation}
which by making use of the definitions collected in the introductory section one can translate into the geometric observables:
\begin{equation}\label{ineq}
|V\theta|\lesssim \frac{3(1-w)}{8}\beta(\nu)\, . 
\end{equation}
Thus the quantum effects do not depend neither on the size of the universe nor  on its expansion rate alone but rather on the specific combination of them both. It is a common belief that the quantum dynamics begins roughly when the energy density of matter hits the Planck scale and the universe is said to enter the Planck era. However, our result does not confirm this assumption. First we note that by virtue of the Friedman equation the energy density of matter content is proportional to the expansion rate squared if the curvature term is neglected \footnote{This is a reasonable assumption, because as the universe approaches the singularity the intrinsic curvature term becomes dominated by the energy density of fluid.}. However, inequality (\ref{ineq}) says that even for the {\it Planck scale} value of expansion rate the universe may still be classical provided it is large enough. On the other hand, a low energy density universe, which is small enough, may undergo a quantum phase.

Let us take a look at our Universe. For simplicity we will assume that the Universe has been filled with radiation from the big bang up to present. Since we are exploring the extension of the classical phase, we may use classical equation of motion. In addition, we set the intrinsic curvature to zero, as the current observations suggest it has not played a significant role in the evolution of the Universe so far. Then $V\theta=V_{0}\theta_{0}\cdot\dfrac{a}{a_0}$, where the subscript $_0$ refers to the present value.
 Let us fix $V_0\sim 10^{81}\,\mathrm{m}^3$ to be the size of the observable Universe today and $\theta_0\sim 10^{-11}c^{-1}\,\mathrm{s}^{-1}$. Let the unit of $\beta(\nu)$ be $\mathrm{m}^2$. Ineq. (\ref{ineq}) is saturated for
\begin{equation}
\frac{a}{a_0}\approx 10^{-62}\beta(\nu)\, . 
\end{equation}
Hence the size of the Universe on the brink of the quantum era cannot be smaller than $V\approx 10^{-186}\beta^{3}(\nu)\,\mathrm{m}^3\approx \beta^{3}(\nu)l_P^3$, while the expansion rate is at least $\theta\approx 10^{113}c^{-1}\,\mathrm{s}^{-1}{\beta^{-2}(\nu)}\approx 10^{58}t_P^{-1}c^{-1}{\beta^{-2}(\nu)}$. Both these values may in principle be deep, deep into the Planck era. In particular we estimate  that the energy density of matter was at that time:
\begin{equation}
\rho\approx 10^{113}\rho_{Pl}{\beta^{-4}(\nu)}\, .
\end{equation}
Now, we may expect that the quantum phase does not start prior to the Planck era and so $\beta(\nu)\lesssim 10^{33}\,\mathrm{m}^2$. Well, this is not strictly a constraint, because $\beta(\nu)$ can be by any order of magnitude larger provided that the Universe is respectively bigger than its observable part. Obviously, the above consideration relies on the simplest possible model of the big bang singularity. Nevertheless, it indicates that the Planck scale argument should be applied more carefully. As shown it is not applicable to the size or the energy density of the universe alone. One may however wonder if in another clock variable frame we would still obtain the same estimation for the domain of validity of classical dynamics. In other words, if inequality (\ref{ineq}) would remain the same upon a different choice of clock, let us say - the size of the universe. At the moment we only notice that the equality (\ref{ineq}) is only weakly sensitive to the value of $w$ within its range.
 
\subsection{Modified Friedman equation} Putting the lower symbol of the Hamiltonian (\ref{lowerH}) back to the constraint equation (\ref{Ham}) we obtain the effective Hamiltonian constraint. In order to get the modified or semiclassical Friedman equation, we need to divide the constraint by lapse $N=q^{2w/(1-w)}$ and by volume $V=q^{\frac{2}{1-w}}$. We get:\footnote{The lapse and volume are functions of physical degrees of freedom and as such are considered as quantum operators. Thus, one might wish to use lower symbols of $N$ and $V$ instead of classical counterparts to perform the division. However, we have already obtained semiclassical expression in (\ref{lowerH}), and we should work with it by treating all the quantities, including $N$ and $V$, classically.}
\[
\alpha(w) p^2q^{-2\frac{1+w}{1-w}} +\frac{3(1-w)^2 a_P^2 K_1(\nu)^2 \left(1+\nu \frac{K_0(\nu)}{K_1(\nu)} \right)}{64 K_0(\nu) K_2(\nu) q^{\frac{4}{1-w}}}+6 \tilde{k} \frac{K_{\mu(w)}(\nu) K_{\mu(w)+1}(\nu)}{K_0(\nu)K_1(\nu)} q^{\frac{-4}{3(1-w)}}\]\[=p_T q^{-2\frac{1+w}{1-w}}\,.
\]
Now, we combine the definitions of $(q,p)$ and $(\tilde{a},\tilde{p})$ to arrive at:
\[
\left( \frac{\dot{a}}{a} \right)^2+c^2a_P^2(1-w)^2\frac{A(\nu)}{V^{2}}+B(\nu)\frac{kc^2}{a^2} =\frac{8\pi G}{3c^2}\rho~,
\]
where \[A(\nu)=\frac{K_1(\nu)^2K(\nu)}{32 K_0(\nu) K_2(\nu)}~,~~B(\nu)=\frac{K_{\mu(w)}(\nu) K_{\mu(w)+1}(\nu)}{K_0(\nu)K_1(\nu)}~.\]
As the result of quantization we obtain two corrections to the Friedman equation. First, the repulsive potential, which depends on the volume. We notice that as the singularity is approached $a\rightarrow 0$, the repulsive potential grows faster ($\sim a^{-6}$) than the density of fluid ($\sim a^{-3(1+w)}$) and therefore at some point the contraction must come to a halt. Second, the curvature becomes dressed by factor $B(\nu)$. This effect could in principle be observed far away from the quantum phase. However, we do not observe the intrinsic curvature neither in the geometry nor in the dynamics of space. Nevertheless, we note that $B(\nu)\approx 1$ for large enough $\nu$. The function $A(\nu) \simeq\frac{\nu}{128}$ for large $\nu$. Assuming $\nu$ is large enough we obtain:
\begin{equation}\label{effF}
\left( \frac{\dot{a}}{a} \right)^2+c^2a_P^2(1-w)^2\frac{\nu}{128}\frac{1}{V^{2}}+\frac{kc^2}{a^2} =\frac{8\pi G}{3c^2}\rho~,
\end{equation}
The form of the repulsive potential featuring in Eq.\,(\ref{effF}) does not depend on the state of fluid, which fills the universe. Therefore, we may conclude that the origin of singularity avoidance is quantum geometrical. Although the potential's coefficient weakly depends on the matter content through $(1-w)^2$, the dependence on $w$ can be absorbed in the definition of $\nu$. The potential provides a kind of hardness to the collapsing space that resists its contraction. As the space contracts the gravitational interaction grows and increases the contraction rate even more. Then, at some point, the potential turns on and makes the contraction slow down until the space comes to a complete halt and rebounds. After the rebound the expansion initially accelerates and the potential turns off shortly after. Afterwards the dynamics becomes classical.

\subsection{Constraint from cosmography}

Szydlowski et al. \cite{szydlowski} have studied cosmographical constraints for the additional source term of the form $\Omega_{mod}(z+1)^6$ in the $\Lambda$CDM model. It corresponds to the geometrical correction in our modified Friedman equation. Interestingly, this kind of modification appears outside our construction in the framework of loop quantum cosmology, brane-world scenarios and others. The authors derived a constraint for the value of $\Omega_{mod}$. The constraint comes from a diverse data set: SNe Ia and radio galaxies sample, baryon oscillation peak (from Sloan Digital Sky Survey) and the so-called CMB shift parameter. The result can be translated into our model's free parameter as:
$$
c^2a_P^2(1-w)^2\frac{\nu}{128 H_0^2 V_0^2}\leqslant 0,26\cdot 10^{-9}
$$
from which we get:
$$
a_P^2(1-w)^2\nu = 2\beta^2(\nu)\lesssim 10^{131}\,\mathrm{m}^4
$$
This constraint is, as expected, milder than the one we got from assuming that the correction comes into a play only after the Universe has entered the Planck era. The authors of \cite{szydlowski} also estimate the constraint from the big bang nucleosynthesis to be:
$$
a_P^2(1-w)^2\nu = 2\beta^2(\nu)\lesssim 10^{120}\,\mathrm{m}^4
$$
which is still a much weaker constraint then the first one that we have arrived at.

\subsection{Conclusion}
The core result of this work is the occurrence of repulsive potential providing a mechanism for the singularity resolution. The potential is quantum geometrical in nature and prevents the space from reaching the singular state. Eventually, the space-time rebounds. The potential is a generic and unexpected feature of our quantization scheme. The coupling constant of the potential depends on the choice of fiducial vector, which is used to generate coherent states. We fully parametrize this freedom with a positive and otherwise arbitrary parameter $\nu$. 

The free parameter is, in our mind, a desired result. First of all, $\nu$ may be used to fit our model into observational constraints.  We may also see the free parameter(s) as ``modeling'' the more fundamental structure of Quantum Gravity. On one hand, our theory is a merger of the principles of quantum mechanics and the dynamics of gravitational field. On the other hand, unlike Weyl quantization, it includes the many ways in which one may do the merger. It encompasses the absence of observational clues of phenomena involving quantized gravity as well as the lack of knowledge of the fundamental principles on which the full Quantum Gravity is perhaps to be achieved one day. Our line of research may contribute to the discovery of this structure. 

Our approach is bottom-up: we begun with quantizing homogenous and isotropic models. Next we will move to more complex ones. The natural extension of this work will be to allow for anisotropic evolution of space, which admits more complex singularity. We also plan to include different choices of clock variable and complement reduced phase space quantization with the Dirac method.

\section*{Acknowledgement}
We thank Nathalie Deruelle for her comment on the manuscript. The work was supported by Polonium program No. 27690XK {\it Gravity and Quantum Cosmology}.

\appendix

\section{Comment on affine quantization of gravity}
\label{A}
The idea of proceeding in quantum  gravity in its Arnowitt-Deser-Misner formulation with an ``affine'' quantization instead of the  Weyl-Heisenberg quantization  was already present in Klauder's  work \cite{klauder_Aslaksen70} and also in \cite{isham84A}. 

More precisely,  one starts from the classical  ``ax + b'' affine algebra with its two generators $q$ (position), $d=pq$ (dilation), built from the usual phase space canonical pair $(q,p)$, $\{q,p\}=1$, and obeying
$\{q,d\}= q$. Then, following the usual canonical quantization procedure, 
$q\mapsto Q$, $p\mapsto P$, with $[Q,P]= i\hbar I$,  one obtains the quantum version of the dilation, $d\mapsto \frac{1}{2} (PQ + QP)$, and the resulting affine commutation rule $[Q,D]= i\hbar Q$. At the difference of the original $Q$ and $P$, the affine operators $Q$ and  $D$ are reducible: there are three inequivalent irreducible self-adjoint representations, $Q>0$, $Q<0$, and $Q=0$.   The quantization of classical observables follows through the usual replacement $f(q,d) \mapsto f(Q,D)$ followed by a symmetrization. Then, a specific family of  affine coherent states $|p,q\rg$ (in Klauder's notation) is built from the unitary action of the affine group
\begin{equation}
\label{affcsfid}
|p,q\rg := e^{ipQ/\hbar}e^{-i \ln(q) D/\hbar}|\tilde\beta\rg\, ,
\end{equation}
on a fiducial vector $|\tilde\beta\rg $ chosen as an extremal weight vector which is a solution of the  first order differential  equation 
\begin{equation}
\label{eqdifffid}
(Q-1 + (i/\tilde\beta)D)|\tilde\beta\rg= 0,
\end{equation}
where $\tilde\beta$ is a free parameter. Note that this equation is the affine counterpart of the $a|0\rg =0$  satisfied by the Gaussian fiducial vector in the case of standard coherent states. 

Given a quantum operator $A$ issued from this scheme, like the Hamiltonian, its mean values or lower symbols $A(p,q) = \lg p,q |A |p,q\rg$ allows to make the classical and quantum theories coexist in a consistent way:  the classical limit of this \textit{enhanced affine quantization} \`a la Klauder is a canonical theory. 

In  \cite{klauder_Aslaksen70} the authors build  a toy model of gravity where  $p>0$ represents the metric with signature constraints and  $q$  represents the Christoffel symbol. In the later work,  Klauder has chosen  $q>0$  for the metric and $ -p$  as the Christoffel symbol.

In a recent paper \cite{fanuel_Zonetti13}, Fanuel and Zonetti follow Klauder's approach to affine quantization to deal with highly symmetric cosmological models.  

J. R. Klauder has also studied affine quantization of the entire gravitational field. As one short and summarizing article look at \cite{klauder11} and references therein. 

Like in our work, this ``enhanced quantization'' provides a natural and non ambiguous regularization of some singularities encountered in Gravity. Actually, one can consider the two approaches as complementary, the Klauder one being still based on the usual canonical procedure, ours being based on integral quantization made possible by a resolution of the identity obeyed by  coherent states. We should add at the credit of our approach that  we leave a huge freedom in the choice of coherent states. In any case, we suspect that the two procedures are physically equivalent (respective quantum observables could differ at higher order in $\hbar$). 

\section{Definition and essential properties of the affine group}
\label{B}
In this appendix, we leave aside all physical dimensions. Consider the  affine transformation on the real line: 
$$
  \mathbb{R} \ni t \rightarrow (a,b)\cdot t := at + b
$$
where the pair dilation-translation parameters $(a, b)$ belongs to $\mathbb{R}_+^* \times \mathbb{R}$.  \\
 The transformations $(a, b)$ form a group with the composition rule:
$$
(a, b) (a', b') = (aa', ab' + b)\,.
$$
The neutral element is $(1,0)$ and the inverse of $(a,b)$ is $(a,b)^{-1}= (1/a, -b/a)$.  This group is called the \emph{affine group} of the real line and denoted by the symbol Aff$_+(\mathbb{R})$. It has a left-action and a right-action on itself. The left-invariant measure is 
$$
\dfrac{\ud a \ud b}{a^2}\,,
$$
whereas the right-invariant one is 
$$
\dfrac{\ud a \ud b}{a}\,. 
$$
In the main text, we use a slightly different realization of the group, namely we perform a transformation of coordinates $(a, b) \rightarrow (q, p)$ such that the measure is of the Lebesgue form, $\ud q \ud p$. One can easily check that this coordinate transformation is
$$
q = 1/a\, , \quad p = b
$$
for the left-action, and
$$
q = \ln(a)\, ,  \quad p = b
$$
for the right-action. This coordinate transformation affects the composition rule as well:
$$
(q, p) (q', p') = \left(q q', \dfrac{p'}{q} + p\right)\, , \quad  (q, p) (q', p') = (q + q', e^{q} p' + p)\, ,
$$
for the left-action and right-action respectively.
\\
In the main text we consider only the left-action. Also, the choice of coordinates $q$ and $p$ is such that they can be thought of as the configuration variable and its conjugated momentum parametrizing the phase space of FLRW cosmology, which is the half-plane $\Pi_+ = \{(q, p) \in \mathbb{R}^2 : q > 0\}$. In particular, since $(q, p) \in \Pi_+$ are canonical variables, the symplectic structure wrt these variables is of the diagonal form, and thus it follows that the measure is Lebesgue. That is why we did the transformation from the mathematically more natural $(a, b)$ to the physically meaningful $(q, p)$.

We already saw how Aff$_+(\mathbb{R})$ acts on $\mathbb{R}$: in term of the new variables,
$$
s \rightarrow (q, p) \cdot s = \dfrac{s}{q} + p\,. 
$$
Let us make explicit  Hilbert spaces, on which the (left-)action of Aff$_+(\mathbb{R})$ is unitarily represented. They are parametrized by $\alpha \in \mathbb{R}$:\footnote{
It has to be said that the quantum theories arising from two Hilbert spaces of this family are unitarily equivalent, so one can effectively choose $\alpha$ at will, as it cannot play any role on the physics. In particular, in the main text the choice is $\alpha = -1$, as this corresponds to the usual  Schr\"odinger representation.
}
$$
\mathcal{H}_\alpha := L^2(\mathbb{R}^*_+, dx/x^{\alpha + 1})\,. 
$$
The fact that these Hilbert spaces are based on the positive half-line, $\mathbb{R}^*_+$, is suggested by the usual situation in QM: the wavefunctions are maps from the classical configuration (in our case, the space where $q$ takes values) to $\mathbb{C}$. The requirement that such functions are square-integrable is also a standard QM requirement, while the weight in the Lebesgue measure is introduced for more generality. In the  concrete study  of physical operators, we conveniently choose $\alpha = -1$, so that the measure becomes the usual one.
\\
The action of the operators $U_\alpha(q, p)$  on $\psi \in \mathcal{H}_\alpha$ is defined as
$$
(U_\alpha(q, p) \psi)(x) = q^{\alpha/2} e^{ipx} \psi(x/q)\,. 
$$
It is easy to check that $U_\alpha(q, p)$ is unitary. Its irreducibility has been shown in \cite{gelnai} (see also \cite{aslaklauder}).
\\
In the main text, we use the action of $U_\alpha(q_0,p_0)$ on a chosen $\psi$ to produce the \emph{affine coherent state}, $| q_0, p_0 \rangle$, peaked on a classical phase space point $(q_0, p_0)$. As a function of $x$ it is  defined  as
$$
\langle x | q_0, p_0 \rangle = (U_\alpha(q_0, p_0) \psi)(x) = q_0^{\alpha/2} e^{ip_0x} \psi(x/q_0)\, . 
$$
Due to the affine group  composition rule, we see that acting with $U_\alpha(q, p)$ on $| q_0, p_0 \rangle$ produces the functions
$$
\langle x | U_\alpha(q, p) | q_0, p_0 \rangle = \langle x | U_\alpha(q, p) U_\alpha(q_0, p_0) | \psi \rangle = \langle x | U_\alpha(q q_0, p_0/q + p) | \psi \rangle = \langle x | q q_0, p_0/q + p \rangle\, , 
$$
i.e.
$$
U_\alpha(q, p) | q_0, p_0 \rangle = | q q_0, p_0/q + p \rangle =: | q', p' \rangle\, .
$$
This shows that the coherent state $| q_0, p_0 \rangle$ transforms covariantly under the action of unitary operator $U_\alpha(q, p)$.
\\
At this point, Schur's lemma applies.
\subsection*{Schur's lemma} \emph{Let $G$ be a group with $U$ its UIR on a vector space $V$. If $M$ is an operator on $V$ such that $U(g) M U(g)^{\dag} = M$ for all $g \in G$, then $M$ is a multiple of the identity $1$ on $V$: $M = c\cdot 1$.}
\\
In our case, it is easy to check that the operator $\int \ud q \ud p | q, p \rangle \langle q, p |$ satisfies the hypotheses of the lemma. Therefore, it follows that
$$
\int \ud q \ud p | q, p \rangle \langle q, p | = c\cdot 1
$$
Combining this operator identity with the  projector $|\psi\rg\lg \psi|$ on a suitably selected unit vector $|\psi\rg$ and taking the trace allow to compute the constant. It is  finite because the UIR is shown to be square integrable.  This is nothing but the statement of the main text that the CS family $\{ |q,p\rg\}$ resolves the identity w.r.t. the measure $\ud q \ud p/c$, and is the starting point for coherent state quantization, and, as well, for continuous wavelet analysis.

\section{Affine quantization for complex fiducial vectors}
\label{C}

In what follows we extend definition (\ref{wavelet}) of wavelets to complex fiducial vectors $\psi_0$, which are rapidly decreasing functions on $\mathbb{R}^*_+$. The quantization of classical functions $f(q,p)$ is again implemented through formula (\ref{quantfaff}). One finds
\begin{equation}
A_{p^2}=P^2+\frac{\ap}{2} L\left(PQ^{-1}+Q^{-1}P\right)+\ap^2\frac{K}{Q^2}
\end{equation}
where we extend definitions of $K$ in (\ref{defK}) and $c_{\alpha}$ in (\ref{cdef}) as follows:
\begin{equation}
K:=\int_0^{\infty}\frac{x \ud x}{c_{-1}}  |\psi_0'(x)|^2~,~~ c_{\alpha} :=  \int_0^{\infty} |\psi_0 (x)|^2\, \frac{\ud x}{x^{2+\alpha}}
\end{equation}
and introduce another constant, which vanishes for real $\psi_0$:
\begin{equation}
L:=i\int_{0}^{+\infty}(\psi_0'\bar{\psi}_0-\bar{\psi}_0'\psi_0)
\end{equation}
We note the extra term in the quantized hamiltonian, which is linear in $P$ and thus may be associated with the expansion of universe. Unlike the repulsive potential, this extra term is present only for complex fiducial vectors. Its role in the singularity resolution will be a subject of a separate investigation. In Appendix \ref{D} we find that it is present also in the Weyl-Heisenberg CS quantization.

\section{Weyl-Heisenberg coherent state quantization}
\label{D}
In what follows we present in fair detail computation of quantum operators for the half-plane observables via the Weyl-Heisenberg CS. In agreement with the affine CS, we find the repulsive potential of the quantum hamiltonian to be a general feature of the employed scheme. In addition, we derive an extra term, which is proportional to expansion and which occurs only for families of CS with complex fiducial vectors.

First, we map the half-plane, canonically parametrized by pair $(q,p)\in \R^{\ast}_+\times\R$ onto the plane with the following canonical parametrization $(\tilde{q},\tilde{p}):=(\ln q,qp)\in \R^2$. In the new variables, we have $q=e^{\q}~,~~p=\p e^{-\q}$ and $p^2=\tilde{p}^2e^{-2\tilde{q}}$. When we take into account the dimension of $q$ and $p$ it becomes physically more meaningful to use $\qm:=\ln \dfrac{q}{\sigma}$ and $\pm:=\dfrac{qp}{\ap}$, where $\sigma$ has the dimension of $q$. Then we have  $q=\sigma e^{\qm}~,~~p=\left(\dfrac{a_P}{\sigma}\right)\pm e^{-\qm}$ and $p^2=\left(\dfrac{a_P}{\sigma}\right)^2\pm^2e^{-2\qm}$.

Any infinite-dimensional UIR of the Weyl-Heisenberg group is characterized by a real number $\lambda\neq 0$ and may be given as
\begin{equation}
U^{\lambda}(\qm,\pm):=e^{-i\lambda\pm\qm/2}e^{i\lambda \pm Q}e^{-i\lambda\qm P}
\end{equation}
acting in $L^2(\mathbb{R}, dx)$  in the following way:
\begin{equation}
U^{\lambda}(\qm,\pm)\phi(x)=e^{i\lambda\pm(x-\qm/2)}\phi(x-\qm)\,.
\end{equation}
Hence $(Q\phi)(x)=x\phi(x)$ and $(P\phi)(x)=-\frac{i}{\lambda}\frac{\partial\phi}{\partial x}(x)$ are the generators of the representation, which satisfy the CCR. We fix $\lambda=1$ and denote the representation simply by $U(\qm,\pm)$. Given a normalized vector $\phi_0\in L^2(\mathbb{R}, dx)$, a continuous family of unit vectors
\begin{equation}
|\qm,\pm\rg = U(\qm,\pm)|\phi_0 \rg\, , \quad \lg x | \qm,\pm\rg = e^{i\pm(x-\qm/2)}\phi_0(x-\qm)
\end{equation}
resolve the unity. They are the Weyl-Heisenberg coherent states in quantum mechanics and Gabor states in time-frequency signal analysis \cite{G2000}. They may be used to implement the quantization of classical function $f(\qm,\pm)$ via\begin{equation}
f\mapsto A_f = \int_{\Pi} \frac{\ud \qm \ud \pm}{2\pi}\, f(\qm,\pm) \, | \qm,\pm\rg\lg \qm,\pm |\, \, . 
\end{equation}
In order to compare quantum operators obtained respectively from the W-H CS and from the affine CS, we introduce the isometry  $T: L^2(\mathbb{R},dx)\ni \phi(x)\mapsto\psi(y):=\frac{\phi(\ln y)}{\sqrt{y}}\in L^2(\mathbb{R}^*_+,dy)$. One easily checks that 
\begin{equation}
Te^{-2x}T^{\dagger}=\frac{1}{y^2}~,~~T\frac{\partial}{\partial x}T^{\dagger}=\sqrt{y}\frac{\partial}{\partial y}\sqrt{y}\, .
\end{equation}
Note that the self-adjoint $-i\partial/\partial x$ acting in $L^2(\R,dx)$ is transformed into the self-adjoint dilation acting in $L^2(\R^\ast_+,dx)$.
\subsection*{The quantized kinetic term}
In what follows we compute $A_{p^2}$:
{
\begin{align}
A_{p^2}=&\left(\frac{a_P}{\sigma}\right)^2\int_{\Pi} \frac{\ud \qm \ud \pm}{2\pi}\, \pm^2e^{-2\qm} \, | \qm,\pm\rg\lg \qm,\pm |\\ \nonumber
=&\left(\frac{a_P}{\sigma}\right)^2\int_{\Pi} \frac{\ud \qm\ud \pm}{2\pi}e^{i\pm(x-x')}\phi_0(x-\qm)\bar{\phi}_0(x'-\qm)\pm^2e^{-2\qm}\\ \nonumber
=&-\left(\frac{a_P}{\sigma}\right)^2\int_{\Pi}  \frac{\ud \qm\ud \pm}{2\pi}\left(e^{i\pm(x-x')}\right)_{,xx}\phi_0(x-\qm)\bar{\phi}_0(x'-\qm)e^{-2\qm}\\ \nonumber
=&-\left(\frac{a_P}{\sigma}\right)^2\frac{\partial^2}{\partial x^2}\left[\int_{\R_+}  \ud \qm\delta(x-x')\phi_0(x-\qm)\bar{\phi}_0(x'-\qm)e^{-2\qm}\right]\\ \nonumber
&+2\left(\frac{a_P}{\sigma}\right)^2\frac{\partial}{\partial x}\left[\int_{\R_+}  \ud \qm\delta(x-x')\phi_0(x-\qm)_{,x}\bar{\phi}_0(x'-\qm)e^{-2\qm}\right]\\ \nonumber
&-\left(\frac{a_P}{\sigma}\right)^2\left[\int_{\R_+}  \ud \qm\delta(x-x')\phi_0(x-\qm)_{,xx}\bar{\phi}_0(x'-\qm)e^{-2\qm}\right]\\ \nonumber
=&\left(\frac{a_P}{\sigma}\right)^2\bigg[-\frac{\partial^2}{\partial x^2}e^{-2x}\delta(x-x')a+2\frac{\partial}{\partial x}e^{-2x}\delta(x-x')b-e^{-2x}\delta(x-x')c\bigg]
\end{align}}
where
\begin{equation}\label{WHcof}
a=\int_{-\infty}^{+\infty} \phi_0(x)\bar{\phi}_0(x)e^{2x},~~b=\int_{-\infty}^{+\infty} \phi_0'(x)\bar{\phi}_0(x)e^{2x},~~c=\int_{-\infty}^{+\infty} \phi_0''(x)\bar{\phi}_0(x)e^{2x}
\end{equation}
Hence,
\begin{equation}
TA_{p^2}T^{\dagger}=\left(\frac{a_P}{\sigma}\right)^2\bigg[-a\frac{\partial^2}{\partial y^2}\delta(y-y')+2(a+b)\frac{1}{y}\frac{\partial}{\partial y}\delta(y-y')+(-\frac{9}{4}a-3b-c)\frac{1}{y^2}\delta(y-y')\bigg]
\end{equation}
To simplify notation we define $\psi_0:=T\phi_0$ and introduce
\begin{equation}
I_{n}:=\int_{0}^{+\infty} y^{n}|\psi_0|^2~,~~J_{n}:=\int_{0}^{+\infty} y^{n}|\psi_0'|^2~,~~K_n:=\int_{0}^{+\infty}y^n(\psi_0'\bar{\psi}_0-\bar{\psi}_0'\psi_0)~,
\end{equation}
where the last quantity is purely imaginary. Now, the coefficients (\ref{WHcof}) read
\begin{eqnarray}
a=I_{2}~,~~b=-I_{2}+\frac{1}{2}K_3+\left(\frac{1}{2}y^{3}|\psi_0(y)|^2\right)\bigg|_{0}^{\infty}\\ \nonumber
c=\frac{13}{4}I_{2}-J_{4}-K_3-\left(y^{3}|\psi_0(y)|^2-\frac{1}{2}y^4(|\psi_0(y)|^2)'-\frac{1}{2}y^4(\psi_0'\bar{\psi}_0-\bar{\psi}_0'\psi_0)\right)\bigg|_{0}^{\infty}
\end{eqnarray} 

\subsection*{The quantized coordinates}
We compute operators for the coordinates in the half-plane $q=\sigma e^{\qm}$ and $p=(\frac{\ap}{\sigma})\pm e^{-\qm}$ as follows:
\begin{equation}A_{q}=\sigma\int_{\Pi} \frac{\ud \qm \ud \pm}{2\pi}\, e^{\qm} \, | \qm,\pm\rg\lg \qm,\pm |=\sigma\int \frac{\ud \qm\ud \pm}{2\pi} e^{i\pm(x-x')}\phi_0(x-\qm)\bar{\phi}_0(x'-\qm)e^{\qm}=\sigma\delta(x-x')I_{-1}e^x\end{equation}
hence
\begin{equation}TA_{q}T^{\dagger}=I_{-1}\sigma y\delta(y-y')~.\end{equation}
The above can be generalized to
\begin{equation}TA_{q^{\alpha}}T^{\dagger}=I_{-\alpha}\sigma^{\alpha} y^{\alpha}\delta(y-y')~.\end{equation}
Also,
\begin{eqnarray} \nonumber A_{p}=\left(\frac{\ap}{\sigma}\right)\int_{\Pi} \frac{\ud \qm \ud \pm}{2\pi}\, \pm e^{-\qm} \, | \qm,\pm\rg\lg \qm,\pm |=\left(\frac{\ap}{\sigma}\right)\int \frac{\ud \qm\ud \pm}{2\pi}   e^{i\pm(x-x')}\phi_0(x-\qm)\bar{\phi}_0(x'-\qm)\pm e^{-\qm}\\ \nonumber
=-i\left(\frac{\ap}{\sigma}\right)\frac{\partial}{\partial x}e^{-x}\delta(x-x')I_{1}+\frac{i}{2}\left(\frac{\ap}{\sigma}\right) \left(-I_{1}+K_2+y^2|\psi_0(y)|^2\bigg|_{0}^{\infty}\right)e^{-x}\delta(x-x')\end{eqnarray} 
hence
\begin{equation}TA_{p}T^{\dagger}=-i\left(\frac{\ap}{\sigma}\right) I_{1}\frac{\partial}{\partial y}\delta(y-y')+\frac{i}{2}\left(\frac{\ap}{\sigma}\right)\left(K_2+y^2|\psi_0(y)|^2\bigg|_{0}^{\infty}\right)\frac{1}{y}\delta(y-y')\end{equation}

\subsection*{The quantized expansion rate}
Similarly, one finds that
\begin{align}\nonumber
TA_{\theta}T^{\dagger}
&=\frac{3}{8}(1-w)\ap\sigma^{-\frac{1+w}{1-w}}\bigg[-\frac{i}{2}I_{\frac{2}{1-w}}\left(y^{-\frac{1+w}{1-w}}\frac{\partial}{\partial y} +\right. \\
&\left. +\frac{\partial}{\partial y}y^{-\frac{1+w}{1-w}}\right)+\frac{i}{2}\left( K_{\frac{3-w}{1-w}}+y^{\frac{3-w}{1-w}}|\psi_0(y)|^2\bigg|_{0}^{\infty}\right)y^{-\frac{2}{1-w}}\bigg]\delta(y-y')
\end{align}

\subsection*{The specification of $\psi_0$}
In what follows we assume $\psi_0$ to be a rapidly decreasing function so that all the boundary terms vanish. Suppose that $\psi_0$ is complex and $K_n\neq 0$. We now define $Q:=\sigma x$ and $P:=-i\frac{\ap}{\sigma}\frac{\partial}{\partial x}$. Then
\begin{equation} A_{p^2}=I_{2}P^2+\frac{i\ap}{2}K_3\left(PQ^{-1}+Q^{-1}P\right)+\ap^2\left(-\frac{10}{4}I_{2}+J_4\right)\frac{1}{Q^2}\end{equation}
is apparently symmetric but its extension to a self-adjoint operator is a separate issue. The basic variables read:
\begin{equation} A_q=I_{-1}Q~,~~A_p=I_{1}P+\frac{i\ap}{2} K_2\frac{1}{Q}\end{equation}
and the expansion rate reads
\begin{equation}A_{\theta}=\frac{3}{16}(1-w)I_{\frac{2}{1-w}}(Q^{-\frac{1+w}{1-w}}P+PQ^{-\frac{1+w}{1-w}})+\frac{3i\ap}{16}(1-w) K_{\frac{3-w}{1-w}}Q^{-\frac{2}{1-w}}\end{equation}
For $\psi_0$ real the above formulas simplify further:
\begin{equation}A_{p^2}=I_{2}P^2+\ap^2\left(-\frac{10}{4}I_{2}+J_4\right)\frac{1}{Q^2}~,~~A_{q}=I_{-1}\hat{Q}~,~~A_{p}=I_{1}\hat{P}\end{equation}
It follows that in order to recover the canonical rule we need $I_{-1}=I_1^{-1}$. Note that the coefficient in front of the potential satisfies:
\begin{equation}
-\frac{10}{4}I_{2}+J_4\geqslant -\frac{1}{4}I_{2},
\end{equation}
which ensures the positivity of $A_{p^2}$.

\end{document}